\documentclass[11pt,final,onecolumn] {IEEEtran}
\usepackage{psfrag}
\usepackage{color}
\usepackage[compress]{cite}
\usepackage[pdftex]{graphicx}
\usepackage{amssymb}
\usepackage[cmex10]{amsmath}
\usepackage{amsthm}
\usepackage{algorithm, algorithmic}
\usepackage{array,enumerate}
\usepackage{multirow}
\usepackage[tight, footnotesize]{subfigure}

\graphicspath{{./figure/}}

\newtheorem{assumption}{Assumption}
\newtheorem{theorem}{Theorem}

\newtheorem{lemma}{Lemma}

\newcommand{\hS}{{\hat{S}}}
\newcommand{\cA}{\mathcal{A}}

\renewcommand{\P}{\mathbb{P}}
\newcommand{\kmax}{k_{\mathrm{max}}}
\newcommand{\C}[2]{{C(#1 \mid #2)}}
\newcommand{\hC}[2]{{\tilde{C}(#1 \mid #2)}}

\newcommand{\ofrac}[1]{{\frac{1}{#1}}}

\newcommand{\ceil}[1]{{\lceil {#1} \rceil}}
\newcommand{\indicator}[1]{{\bf 1}_{\{#1\}}}
\newcommand{\tc}[1]{^{(#1)}}
\newcommand{\parent}[1]{\mathrm{pa}(#1)}
\newcommand{\children}[1]{\mathrm{ch}(#1)}

\begin{document}
\title{Identifying Infection Sources and Regions in Large Networks}

\author{Wuqiong~Luo,~\IEEEmembership{Student Member,~IEEE}, Wee~Peng~Tay,~\IEEEmembership{Member,~IEEE} and Mei~Leng,~\IEEEmembership{Member,~IEEE}
\thanks{
This research was supported by the MOE AcRF Tier 1 Grant M52040000. W. Luo, W.~P. Tay and M. Leng are with the Nanyang Technological University, Singapore. E-mail: \texttt{wluo1@e.ntu.edu.sg, \{wptay,lengmei\}@ntu.edu.sg}
}
}

\maketitle

\begin{abstract}
Identifying the infection sources in a network, including the index cases that introduce a contagious disease into a population network,  the servers that inject a computer virus into a computer network, or the individuals who started a rumor in a social network, plays a critical role in limiting the damage caused by the infection through timely quarantine of the sources. We consider the problem of estimating the infection sources and the infection regions (subsets of nodes infected by each source) in a network, based only on knowledge of which nodes are infected and their connections, and when the number of sources is unknown a priori. We derive estimators for the infection sources and their infection regions based on approximations of the infection sequences count. We prove that if there are at most two infection sources in a geometric tree, our estimator identifies the true source or sources with probability going to one as the number of infected nodes increases. When there are more than two infection sources, and when the maximum possible number of infection sources is known, we propose an algorithm with quadratic complexity to estimate the actual number and identities of the infection sources. Simulations on various kinds of networks, including tree networks, small-world networks and real world power grid networks, and tests on two real data sets are provided to verify the performance of our estimators.
\end{abstract}


\section{Introduction}\label{sec:Introduction}
With rapid urbanization and advancements in transportation technologies, the world has become more interconnected. A contagious disease like Severe Acute Respiratory Syndrome (SARS) can spread quickly through a population and lead to an epidemic \cite{Goh2006}. It is crucial to quickly identify the index cases of a contagious disease since it allows us to study the causes, and hence facilitate the search for antiviral drugs and efficacious therapies. Moreover, by inferring the the set of individuals infected by each source, potential containment policies can be formulated to prevent further spreading of the disease due to new index cases \cite{Scoglio2010,Newman2004}. In a similar vein, a computer virus on a few servers of a computer network can quickly spread to other servers or computers in the network. Without prompt identification and isolation of the source servers, significant damage can result \cite{Kephart1991,Han2008}. Identifying the servers in the network that are first infected also allows us to detect the latent points of weaknesses in the computer network so that preventive measures can be taken to enhance the protection at these points. The source identification problem also arises in the study of rumor spreading in a social network. A rumor started by a few individuals can spread quickly through the underlying social network \cite{Weng2010, Bakshy2011, Lim2011, Akritidis2011}. In many cases, we are interested to find the sources of the rumor. For example, law enforcement agencies may be interested in identifying the perpetrators who fabricate false information to manipulate the market prices of certain stocks.

We can model all the above examples as an infection spreading in a network of nodes. In a population network, the infection is the disease that is transmitted between individuals. In the example of a computer virus spreading in a network, the infection is the computer virus, while for the case of a rumor spreading in a social network, the infection is the rumor.  We consider the problem of estimating the infection sources in a network of infected nodes. We are interested in the scenario where the only given information is the set of infected nodes and their connections. This is because typically, complete data about the infection spreading process, like the first times when the infection is detected at each node, is not available. Even when such detection times are available, the naive method of declaring the first detected node in the network as the sole infection source is often incorrect, as the infection may have a random dormant period, the length of which varies from node to node. For example, the spreading of a disease in a population with individuals having varying degrees of resistance, and hence exhibiting symptoms not necessarily in the order in which they are infected, presents such a problem. Our goal is to construct estimators for both the infection sources and their infection regions, i.e., the subset of nodes likely to be infected by each source, when the number and locations of the sources are unknown a priori.

\subsection{Related Works}\label{subsection:Related_works}
Existing works related to infection spreading in a network have primarily focused on the parameters of the diffusion process such as the outbreak thresholds and the effect of network structures \cite{Moore2000, Newman2002, Oneill2002, Ganesh2005}. Little work has been done on identifying the infection sources. Our aim is to identify a set of nodes most likely to be the infection sources after the infection has spread for some time. This formulation is of interest in various practical scenarios, including the spreading of a new disease in a population network. By identifying the initial infectious sources, we can focus scarce resources like DNA testing on a small select group of patients instead of on the whole population. Other examples include identifying the initial entry points of a computer virus into a computer network, and the initiators of a rumor in a social network.

The case where there is a single infection source has been studied in \cite{Shah2011}. Based only on the knowledge of which nodes are infected and the underlying network structure, an estimator based on the linear extensions count of a poset or the number of infection sequences (cf.\ Section \ref{sec:problem_formulation}) was derived to identify the most likely infection source. It was shown in \cite{Shah2011} that finding a single infection source is a \#P-complete problem even in the case where the infection is relatively simple, with infection from an infected node being equally likely to be transmitted to any of its neighbors at each time step. This simple infection model is based on the classical \emph{susceptible-infected} (SI) model \cite{Bailey1975}, which has been widely used in modeling viral epidemics \cite{Allen1994, Barthelemy2004, Yan2005, Zhou2005, Zhou2006, Tang2011}. An algorithm for evaluating the single source estimator was proposed in \cite{Shah2011}, and it was shown to have complexity\footnote{A function $f(n)=O(g(n))$ if $f(n) \leq c g(n)$ for some constant $c$ and for all $n$ sufficiently large.} $O(n)$ for tree networks, where $n$ is the total number of infected nodes. Furthermore, it was shown that this estimator performs well in a very general class of tree networks known as the geometric trees (cf.\ Section \ref{subsection:GeometricTrees}), and identifies the infection source with probability going to one as $n$ increases.

In many applications, there may be more than one infection source in the network. For example,  an infectious disease may be brought into a country through multiple individuals. Multiple individuals may collude in spreading a rumor or malicious piece of information in a social network. In this paper, we investigate the case where there may be multiple infection sources, and when the number of infection sources is unknown a priori. We also consider the problem of estimating the infection region of each source, and show that a direct application of the algorithm in \cite{Shah2011} performs significantly worse than our proposed algorithms if there are more than one infection sources. We also note that \cite{Shah2011} provides theoretical performance measures for several classes of tree networks, which we are unable to do here except for the class of geometric trees, because of the greater complexity of our proposed algorithms. Instead, we provide simulation results to verify the performance of our algorithms.

A related problem is the detection and localization of diffusive sources using wireless sensor networks \cite{Zhao2007, Fox2007, Tay2009, Tay2008, Bianchi2011, Aldalahmeh2011}. The diffusion models used under this framework are based on spatio-temporal diffusion models \cite{Zhao2007} or state-space models with linear dynamics \cite{Fox2007}, where information like the physical positions of sensors are known. There is no natural translation of the source detection and localization problem in a sensor network to other networks like a computer network, without performing discretization and introducing a combinatorial aspect to the problem, as is done in \cite{Kempe2003} and \cite{Kimura2010}. Similarly, inference of viral epidemic processes in populations has been studied in \cite{Bailey1975, Moore2000, Oneill2002}, where various features related to the propagation of a viral epidemic, such as the rates of infection and the length of latency periods are investigated. These works' focus is on specific viral infection processes with assumptions that do not naturally hold for infection processes in other networks. Moreover, there is little work on determining the sources or index cases of a disease.

On the other hand, the infection source estimation algorithms we consider in this paper can be useful in applications like pollution source localization, where we are limited to inexpensive sensors capable only of detecting the presence or absence of a pollutant, and the identities of its neighbors. In this case, spatio-temporal diffusion models are not applicable as we only have knowledge of which nodes are ``infected''. The algorithms we study in this paper are also applicable to inferring infection sources in viral epidemics, when little information about the epidemic propagation characteristics is available.

\subsection{Our Contributions}\label{subsection:Contributions}
In this paper, we consider the estimation of multiple infection sources when the number of infection sources is unknown a priori. We adopt the same SI diffusion model as in \cite{Shah2011}, as this has been widely used to model various infection spreading processes \cite{Allen1994, Barthelemy2004, Yan2005, Zhou2005, Zhou2006, Tang2011}. The results of this work are applicable to scenarios where the infection spreads in an approximately homogeneous way, with infections happening independently. Examples include the spreading of a new disease in a human population, where nobody has yet developed any immunity to the disease. A novel computer virus attacking a network can also be modeled using a homogeneous spreading process. On the other hand, our model is highly simplistic and does not model many other spreading processes of practical interest. However, as alluded to earlier, finding the infection sources in this simple model is already very challenging. The focus of our work is not on modeling infection processes. Rather, by restricting our analysis to the simplest homogeneous exponential spreading model, we hope to gain insights into identifying multiple infection sources in real networks. We show that unlike the single source estimation problem, the multiple source estimation problem is much more complex and cannot be solved exactly even for regular trees. Our main contributions are the following.
\begin{enumerate}[(i)]
\item\label{it:con1} For the case of a tree network, and when it is known that there are two infection sources, we derive an estimator for the infection sources based on the infection sequences count. The estimator can be calculated in $O(n^2)$ time complexity, where $n$ is the number of infected nodes.
\item When there are at most two infection sources that are at least two hops apart, we derive an estimator for the class of geometric trees based on approximations of the estimator in \eqref{it:con1}, and we show that our estimator correctly estimates the number of infection sources and correctly identifies the source nodes, with probability going to one as the number of infected nodes increases.
	\item We derive an estimator for the infection regions of every infection source under a simplifying technical condition.
	\item For general graphs, when there are at most $\kmax$ infection sources, we provide an estimation procedure for the infection sources and infection regions. Simulations suggest that on average, our estimators are within a few hops of the true infection sources in the infection graph.\footnote{In general, we do not know the whole underlying network, but rather the subgraph of infected nodes. For example, in the case of a contagious disease spreading in a population, we only perform contact tracing on the patients to construct the connections among them. From our simulation studies, the infection graph typically has an average diameter of more than 27 hops even though the underlying network's diameter is much smaller.}
	\item We test our estimators on real data in Section \ref{subsec:experiments}. The first test is based on real contact tracing data of a patient cluster during the SARS outbreak in Singapore in 2003. Our estimator correctly identifies the number of index cases for the cluster to be one and successfully finds this index case. The second test considers the Arizona-Southern California cascading power outages in 2011. Our estimator correctly identifies the number of outage sources for the main affected power network to be two, and the distance between our estimators and the real sources are within 1 hop. These tests suggest that our estimator has reasonable performance in some applications even though we have adopted a simplistic infection model.
\end{enumerate}

The rest of the paper is organized as follows. In Section \ref{sec:problem_formulation}, we present the system model and problem formulation. In Section \ref{sec:source_estimation_trees}, we derive estimators for infection sources and regions for tree networks, and present algorithms to evaluate them. We also show asymptotic results for geometric tree networks. We discuss estimation algorithms for general graphs in Section \ref{sec:source_estimation_general}. In Section \ref{sec:simulation_results}, we present simulations and tests on real data to verify the performance of our proposed estimators. Finally we conclude and summarize in Section \ref{sec:conclusion}.

\section{Problem Formulation}\label{sec:problem_formulation}
In this section, we describe our model and assumptions, introduce some notations, and present some preliminary results. Consider an undirected graph $G = (V,E)$, where $V$ is the set of nodes and $E$ is the set of edges. If there is an edge connecting two nodes, we say that they are neighbors. The neighborhood $\mathcal{N}_G(v)$ of a node $v$ is the set of all neighbors of $v$ in $G$. The length of the shortest path between $u$ and $v$ is denoted as $d(u,v)$. In a computer network, the graph $G$ models the interconnections between computers in the network. In the example of a population or a social network, $V$ is the set of individuals, while an edge in $E$ represents a relationship between two individuals. We define an \textbf{infection} to be a property that a node in $G$ possesses, and can be transmitted to a neighboring node. When a node has an infection, we say that it is infected. The neighbors of an infected node is said to be susceptible. We assume the susceptible-infected model \cite{Bailey1975}, where once a node has been infected, it will not lose its infection. We adopt the same infection spreading process as in \cite{Shah2011}, where the time taken for an infected node to infect a susceptible neighbor is exponentially distributed with rate $1$. All infections are independent of each other. Therefore, if a susceptible node has more than one infected neighbors and subsequently becomes infected, its infection is transmitted by one of its infected neighbors, chosen uniformly at random. For mathematical convenience, we also assume that $G$ is large so that boundary effects can be ignored in our analysis.

Suppose that at time $0$, there are $k \geq 1$ nodes in the infected node set $S^*=\{s_1,\ldots,s_k\} \subset V$. These are the \textbf{infection sources} from which all other nodes get infected. Suppose that after the infection process has run for some time, and $n$ nodes are observed to be infected. Typically, $n$ is much larger than $k$. These nodes form an \textbf{infection graph} $G_n = (V_n, E_n)$, which is a subgraph of $G$. Let $\cA^*_n = \cup_{i=1}^k A_{n,i}$ be a partition of the infected nodes $V_n$ so that $A_{n,i} \cap A_{n,j} = \emptyset$ for $i\ne j$, with each partition $A_{n,i}$ being connected in $G_n$, and consisting of the nodes whose infection can be traced back to the source node $s_i$. The set $A_{n,i}$ is called the \textbf{infection region} of $s_i$, and we say that $\cA^*_n$ is the \textbf{infection partition}. Given $G_n$, our objective is to infer the sources of infection $S^*$ and to estimate $\cA^*_n$. In addition, if we do not have prior knowledge of the number of infection sources $k$, we also aim to infer the number of infection sources. Without loss of generality, we assume that $G_n$ is connected, otherwise the same estimation procedure can be performed on each of the components of the graph. We also assume that there are at most $\kmax$ infection sources, i.e., the number of infection sources $k \leq \kmax$. From a practical point of view, if two infection sources are close to each other, we can ignore either one of them and treat the infection as spreading from a single source. Therefore, we are interested in cases where the infection sources are separated by a minimum distance. These assumptions are summarized in the following.
\begin{assumption}\label{assumption:infectiongraph} The number of infection sources is at most $\kmax$, and the infection graph $G_n$ is connected.
\end{assumption}
\begin{assumption}\label{assumption:source_separation}
For all $s_i,s_j \in S^*$, the length of the shortest path between them $d(s_i, s_j) \geq \tau$, where $\tau$ is a constant greater than 1.
\end{assumption}
\begin{assumption}\label{assumption:boundeddeg_tree}
Every node in $G$ has bounded degree, with $d_*$ being the maximum node degree.
\end{assumption}

Suppose that our priors for $S^*$ and $\cA^*_n$ are uniform over all possible realizations, and let $\P$ be the probability measure of the infection process. We seek $S$ and $\cA_n$ that maximize the posterior probability of $S^*$ and $\cA^*_n$ given $G_n$,
\begin{align}
\P(S^* = S,\cA^*_n = \cA_n \mid G_n) \propto P(G_n \mid S)P(\cA_n \mid S, G_n), \label{posterior}
\end{align}
where $P(G_n \mid S)$ is the probability of observing $G_n$ if $S$ is the set of infection sources, and $P(\cA_n \mid S, G_n)$ is the probability that $\cA^*_n=\cA_n$ conditioned on $S$ being the infection source set and the infection graph being $G_n$.

\begin{figure}[!htb] 
  \centering
  \psfrag{a}[][][0.9][0]{$s_1$}
  \psfrag{b}[][][0.9][0]{$s_2$}
  \psfrag{c}[][][0.8][0]{$G_n$}
  \psfrag{d}[][][0.8][0]{$G$}
  \psfrag{e}[][][0.9][0]{$u_1$}
  \psfrag{f}[][][0.9][0]{$u_2$}
  \psfrag{g}[][][0.9][0]{$u_3$}
  \psfrag{h}[][][0.9][0]{$u_4$}
  \psfrag{i}[][][0.9][0]{$u_5$}
  \includegraphics[width=0.35\textwidth]{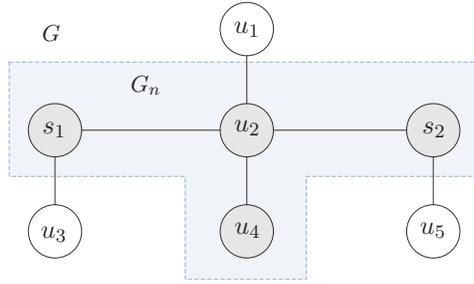}
  \caption{Example of an infection sequence. The shaded nodes are the infected nodes which form the infection graph $G_n$. Infection sources are $S = \{s_1,s_2\}$. The sequence $(u_2,u_4)$ is an infection sequence, but $(u_4,u_2)$ is not. The probability of the infection sequence $\sigma=(u_2,u_4)$ is then given by $P(\sigma\mid S) = \frac{2}{4} \times \frac{1}{4}=\frac{1}{8}$. The first fraction $\frac{2}{4}$ is obtained by observing that when only $s_1$ and $s_2$ are infected, there are four edges $(s_1,u_2)$, $(s_1,u_3)$, $(s_2,u_2)$, and $(s_2,u_5)$ for the infection to spread. The infection is equally likely to spread along any of these four edges, out of which two results in the infection of node $u_2$. After $u_2$ is infected, there are 4 edges over which the infection can spread and this corresponds to the fraction $\frac{1}{4}$.}
  \label{fig:constrain}
\end{figure}

For any source set $S$, let an \textbf{infection sequence} $\sigma = (\sigma_1,\ldots,\sigma_{n-k})$ be a sequence of the nodes in $G_n$, excluding the the $k$ source nodes in $S$, arranged in ascending order of their infection times (note that with probability one, no two infection times are the same). For any sequence to be an infection sequence, a necessary and sufficient condition is that any infected node $\sigma_i$, $i=1,\ldots,n-k$, has a neighbor in $S\cup\{\sigma_1,\ldots,\sigma_{i-1}\}$. We call this the \textit{infection sequence property}. An example is shown in Figure \ref{fig:constrain}. Let $\Omega(G_n,S)$ be the set of infection sequences for an infection graph $G_n$ and source set $S$, and let $\C{S}{G_n} = |\Omega(G_n,S)|$ be the number of infection sequences. We have
\begin{align}\label{eqn:posterior_G}
P(G_n \mid S) = \sum_{\sigma\in\Omega(G_n,S)} P(\sigma \mid S),
\end{align}
where $P(\sigma \mid S)$ is the probability of obtaining the infection sequence $\sigma$ conditioned on $S$ being the infection sources.

Evaluating the expression \eqref{eqn:posterior_G} and maximizing \eqref{posterior} for a general $G_n$ is a computationally hard problem as it involves combinatorial quantities.
As shown in \cite{Shah2011}, if $G$ is a regular tree and $|S|=1$, $P(G_n \mid S)$ is proportional to $|\Omega(G_n,S)|$, which is equivalent to the number of linear extensions of a poset. It is known that evaluating the linear extensions count is a \#P-complete problem \cite{Brightwell1991}. As such, we will make a series of approximations to simplify the problem, and present numerical results in Section \ref{sec:simulation_results} to verify our algorithms. The first approximation we make is to evaluate the estimators
\begin{align}
\hS &= \arg\max_{\substack{S \subset V_n \\ |S| \leq \kmax}} P(G_n \mid S), \label{estimator_S}\\
\hat{\cA}_n(\hS) &= \arg\max_{\cA_n} P(\cA_n \mid \hS, G_n), \label{estimator_A}
\end{align}
instead of the exact maximum a posteriori (MAP) estimators for \eqref{posterior}. Even with this approximation, the optimal estimators are difficult to compute exactly, and may not be unique in general. Therefore, our goal is to design algorithms that are approximately optimal but computationally efficient. In Section \ref{sec:source_estimation_trees}, we make further approximations and design algorithms to evaluate the estimators $\hS$ and $\hat{\cA}_n(\hS)$ when $G$ is a tree. In Section \ref{sec:source_estimation_general}, we consider the case when $G$ is a general graph. For the reader's convenience, we summarize some notations commonly used in this paper in Table \ref{table:notation}. Several notations have been introduced previously, while we formally define the remaining ones in the sequel where they first appear.
\begin{table}[!t]
  \centering
  \caption{Summary of notations used.}\label{table:notation}
    \begin{tabular}{|c|l|}
    \hline
    \textbf{Symbol} & \textbf{Definition}  \\  \hline
    $G$ & underlying network \\ \hline
    $d(u,v)$ & length of the shortest path between $u$ and $v$  \\  \hline
    $\mathcal{N}_G(u)$ & set of neighbors of $u$ in $G$\\ \hline
    $\deg_{G}(u)$ & number of neighbors of node $u$ in $G$ \\  \hline
	$G_n$ &  infection graph with $n$ infected nodes \\ \hline
	$S^*$ & infection sources \\ \hline
    $\cA^*_n $ & infection partition of an infection graph $G_n$ \\  \hline
    $A_{n,i}$ & infection region of an infection source $s_i$ \\  \hline
     \multicolumn{1}{|c|}{\multirow{2}{*}{$\Omega(G_n,S)$}} & set of infection sequences for an infection graph $G_n$ \\
     \multicolumn{1}{|c|}{\multirow{1}{*}{}} & and source set $S$ \\  \hline
    $\C{S}{G_n}$ & $=|\Omega(G_n,S)|$ \\ \hline \hline
    \textbf{Symbol} & \textbf{Definition} (defined implicitly w.r.t.\ $G_n$)  \\  \hline
    $\rho(u,v)$ & path between $u$ and $v$ in the infection graph $G_n$ \\ \hline
    $T_v(S)$ & tree in $G_n$, rooted at $v$ w.r.t. source set $S$ \\ \hline
    $T_M(S)$ & $= \cup_{v\in M} T_v(S)$, where $M$ is a subset of nodes \\ \hline
    $I_i(\xi; S)$ & $= \sum_{j\leq i} |T_{\xi_j}(S)|$, where $\xi$ is a sequence of nodes \\ \hline
    \multicolumn{1}{|c|}{\multirow{2}{*}{$I^*_i(s_1,s_2)$}} & total number of nodes in the $i$ biggest trees \\
    \multicolumn{1}{|c|}{\multirow{2}{*}{}} & in $\{T_{u}(s_1,s_2) : u \in \rho(s_1,s_2)\}$ \\ \hline
    \end{tabular}
\end{table}

\section{Identifying Infection Sources and Regions for Trees}\label{sec:source_estimation_trees}

In this section, we consider the problem of estimating the infection sources and regions when the underlying network $G$ is a tree. We first derive an estimator for the infection partition in \eqref{estimator_A}, given any source node set $S$ and $G_n$. Then, we derive an estimator based on the number of infection sequences. Next, we consider the case where there are two infection sources, propose approximations that allow us to compute the estimator with reasonable complexity, and show that our proposed estimator works well in an asymptotically large geometric tree under some simplifying assumptions. In most practical applications, the number of infection sources is not known a priori. We present a heuristic algorithm for general trees to estimate the infection sources when the number of infection sources is unknown, but bounded by $k_{\max}$.

\subsection{Infection Partition with Multiple Sources}\label{subsection:InfectionRegions}

In this section, we derive an approximate infection partition estimator for \eqref{estimator_A} given any infection source set $S$. This estimator is exact under a simplifying technical condition given in Theorem \ref{theorem:A} below, the proof of which is provided in Appendix \ref{appendix:theorem:A}.

\begin{theorem}\label{theorem:A}
Suppose that $G$ is a tree with infection sources $S$, and $H_n$ is the subgraph of $G_n$ consisting of all paths between any pair of nodes in $S$. If any two paths in $H_n$ do not intersect except possibly at nodes in $S$, then the optimal estimator $\hat{\cA}_n(S)$ for the infection partition is a Voronoi partition of the graph $G_n$, where the centers of the partitions are the infection sources $S$.
\end{theorem}

A Voronoi partition may not produce the optimal estimator for the infection partition in a general infection graph. However, it is intuitively appealing as nodes closer to a particular source are more likely to be infected by that source. For simplicity, we will henceforth use the Voronoi partition of the infection graph $G_n$ as an estimator for $\cA^*_n$, and present simulation results in Section \ref{sec:simulation_results} to verify its performance. We will also see in Section \ref{subsec:UnknownSourceNumber} that this approximation allows us to design an infection source estimation algorithm with low complexity.

\subsection{Estimation of Infection Sources}\label{subsection:InfectionSources}
We now consider the problem of estimating the set of infection sources $S^*$. When $|S^*|=1$, our estimation problem reduces to that in \cite{Shah2011}, which considers only the single source infection problem. In the following, we introduce some notations, and briefly review some relevant results from \cite{Shah2011}.

\begin{figure}[!tb] 
  \centering
  \psfrag{a}[][][0.8][0]{$s_1$}
  \psfrag{b}[][][0.8][0]{$s_2$}
  \psfrag{c}[][][0.8][0]{$u_1$}
  \psfrag{d}[][][0.8][0]{$u_m$}
  \psfrag{e}[][][0.8][0]{$n_1$}
  \psfrag{f}[][][0.8][0]{$n_2$}
  \psfrag{g}[][][0.8][0]{$n_i$}
  \psfrag{h}[][][0.8][0]{$n_{i+1}$}
  \psfrag{i}[][][0.8][0]{$n_{i+2}$}
  \psfrag{j}[][][0.8][0]{$n_k$}
  \psfrag{k}[][][0.8][0]{$T_{n_2}(S)$}
  \psfrag{l}[][][0.8][0]{$T_{u_1}(S)$}
  \psfrag{m}[][][0.8][0]{$T_{\rho(u_1,u_m)}(S)$}
  \psfrag{n}[][][0.8][0]{$H_n$}
  \includegraphics[width=0.6\textwidth]{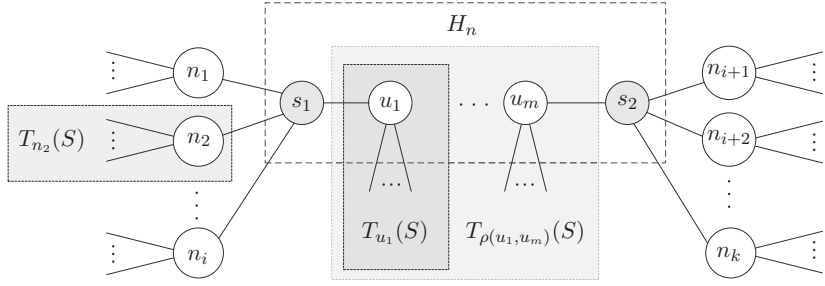}
  \caption{A sample infection graph with $S=\{s_1, s_2\}$}
  \label{fig:two_sources_network}
\end{figure}

A path between any two nodes $u$ and $v$ in the tree $G_n$ is denoted as $\rho(u,v)$. For any set of nodes $S$ in $G_n$, consider the connected subgraph $H_n \subset G_n$ consisting of all paths between any pair of nodes in $S$. Treat this subgraph as a ``super'' node, with the tree $G_n$ rooted at this ``super'' node. For any node $v \in G_n\backslash H_n$, we define $T_v(S)$ to be the tree rooted at $v$ with the path from $v$ to $H_n$ removed. For $v \in H_n$, we define $T_v(S)$ to be the tree rooted at $v$ so that all edges between $v$ and its neighbors in $H_n$ are removed.\footnote{As $T_v(S)$ is defined on $G_n$, its notation should include $G_n$. However, in order to avoid cluttered expressions, we drop $G_n$ in our notations. Confusion will be avoided through the context in which these trees are referenced.} We say that $T_v(S)$ is the tree rooted at $v$ with respect to (w.r.t.) $S$. For any subset of nodes $M \subset G_n$, we let $T_M(S) = \cup_{v\in M} T_v(S)$. An illustration of these definitions is shown in Figure \ref{fig:two_sources_network}. If $S=\{s_1,\ldots,s_k\}$, we will sometimes use the notation $T_v(s_1,\ldots,s_k)$ instead.

Recall that  $\C{S}{G_n}$ is the number of infection sequences if $S$ is the infection source set. If there is a single infection source node $S=\{s\}$, and $G$ is a regular tree where each node has the same degree, it is shown in \cite{Shah2011} that the MAP estimator for the infection source is obtained by evaluating $\hS = \arg\max_{v\in G_n} \C{v}{G_n}$, which seeks to maximize $\C{v}{G_n}$ over all nodes. Therefore, it has been suggested that $\C{v}{G_n}$ can be used as the infection source estimator for general trees. The following result is provided in \cite{Shah2011}.

\begin{lemma}\label{lemma:C_onesource}
Suppose that $G_n$ is a tree. For any node $s\in G_n$, we have
\begin{align}\label{eqn:C_onesource}
\C{s}{G_n} = n! \prod_{u\in G_n} |T_u(s)|^{-1}.
\end{align}
\end{lemma}
We observe that each term $|T_u(s)|$ in the product on the right hand side (R.H.S.) of \eqref{eqn:C_onesource} is the number of nodes in the sub-tree $T_u(s)$ (and which appears when we account for the number of permutations of these nodes). We can think of the terms in the product being ordered according to the infection spreading sequence, i.e., each time we reach a particular node $u$, we include terms corresponding to the nodes $u$ can potentially infect. This interpretation is useful in helping us understand the characterization in Lemma \ref{lemma:C_twosources} for the case where there are two infection sources.

To compute $\C{v}{G_n}$, an $O(n)$ algorithm based on Lemma \ref{lemma:C_onesource} was provided in \cite{Shah2011}. We call this algorithm the Single Source Estimation (SSE) algorithm. We refer the reader to \cite{Shah2011} for details about the implementation of the algorithm. Although finding $\hS$ by maximizing $\C{s}{G_n}$ is exact only for regular trees, it was shown in \cite{Shah2011} that this estimator has good performance for other classes of trees. In particular, if $G$ is a geometric tree (cf.\ Section \ref{subsection:GeometricTrees}), then the probability, conditioned on $S^*=\{s\}$, of correctly identifying $s$ using $\C{s}{G_n}$ goes to one as $n\to\infty$. Inspired by this result, we propose estimators based on quantities related to $\C{S}{G_n}$ for cases where $|S^*| > 1$. In the following, we first discuss the case where $|S^*|=2$, and extend the results to the general case where $|S^*|$ is unknown in Section \ref{subsec:UnknownSourceNumber}. We then numerically compare our proposed algorithms with a modified SSE algorithm adapted for finding multiple sources in Section \ref{sec:simulation_results}.

\subsection{Two Infection Sources} \label{subsection:tree_twosources}
\begin{figure}[!tb] 
  \centering
  \psfrag{a}[][][0.8][0]{$s_1$}
  \psfrag{b}[][][0.8][0]{$s_2$}
  \psfrag{c}[][][0.8][0]{$u_1$}
  \psfrag{d}[][][0.8][0]{$u_2$}
  \psfrag{e}[][][0.8][0]{$u_3$}
  \includegraphics[width=0.4\textwidth]{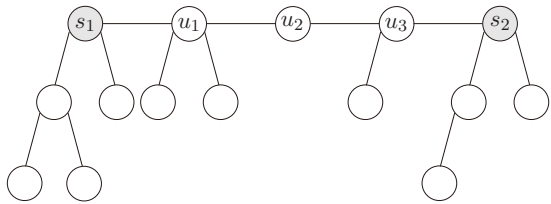}
  \caption{A sample infection graph with $S=\{s_1, s_2\}$. Given an infection sequence $\sigma = (u_3, u_1, u_2)  \in \Omega(\rho(s_1,s_2), \{s_1,s_2\})$, we can find the corresponding reverse infection sequence $\xi =(u_2,u_1,u_3)$. We have $I_1(\xi;s_1,s_2)=|T_{u_2}(s_1,s_2)|=1$, $I_2(\xi;s_1,s_2)=|T_{u_2}(s_1,s_2)|+|T_{u_1}(s_1,s_2)|=4$, $I_3(\xi;s_1,s_2)=|T_{u_2}(s_1,s_2)|+|T_{u_1}(s_1,s_2)|+|T_{u_3}(s_1,s_2)|=6$.}
  \label{fig:sample_network}
\end{figure}

In this section, we assume that there are two infection sources $S=\{s_1,s_2\}$. Given two nodes $u$ and $v$ in $G_n$, suppose that $|\rho(u,v)| = m$. For any permutation $\xi = (\xi_1,\ldots,\xi_m)$ of the nodes in $\rho(u,v)$, let
\begin{align}\label{eqn:I}
I_i(\xi; s_1,s_2) = \sum_{j\leq i} |T_{\xi_j}(s_1,s_2)|
\end{align}
be the total number of nodes in the trees rooted at the first $i$ nodes in the permutation $\xi$. We have the following characterization for $\C{s_1,s_2}{G_n}$, whose proof is given in Appendix \ref{appendix:lemma:C_twosources}.

\begin{lemma}\label{lemma:C_twosources}
Suppose that $G_n$ is a tree. Consider any two nodes $s_1$ and $s_2$ in $G_n$, and suppose that $\rho(s_1,s_2) = (s_1,u_1,\ldots,u_m,s_2)$. We have
\begin{align}\label{eqn:C_two}
\C{s_1,s_2}{G_n} = (n-2)!\cdot q(u_1,u_m; s_1,s_2)\cdot \prod_{u\in G_n\backslash \rho(s_1,s_2)} |T_u(s_1,s_2)|^{-1},
\end{align}
where for $1\leq i\leq j \leq m$, $q(u_i,u_j;s_1,s_2)$ satisfies the following recursive relationship
\begin{align}
& q(u_i,u_j;s_1,s_2) = |T_{\rho(u_i,u_j)}(s_1,s_2)|^{-1}\left(q(u_{i+1},u_j;s_1,s_2) + q(u_i,u_{j-1};s_1,s_2)\right) \ \textrm{for $i<j$,}\label{eqn:q_recurs}
\end{align}
with $q(v,v;s_1,s_2) = |T_v(s_1,s_2)|^{-1}$ for all $v\in \rho(u_1,u_m)$.
Furthermore, we have
\begin{align}\label{eqn:q}
q(u_1,u_m; s_1,s_2) = \sum_{\xi\in\Gamma(u_1,u_m)} \prod_{i=1}^m I_i(\xi;s_1,s_2)^{-1},
\end{align}
and $\Gamma(u_1,u_m)$ is the set of all permutations $\xi = (\xi_1,\ldots,\xi_m)$ of nodes in $\rho(u_1,u_m)$ such that $(\xi_m,\ldots,\xi_1)$ is an infection sequence starting from $s_1$ and $s_2$ and resulting in $\rho(s_1,s_2)$.
\end{lemma}

The characterization for $\C{s_1,s_2}{G_n}$ is similar to that for the single source case in \eqref{eqn:C_onesource}, except for the additional $q(u_1,u_m;s_1,s_2)$ term. We first clarify the meaning of $\Gamma(u_1,u_m)$. Given any infection sequence $\sigma$ that starts with $\{s_1, s_2\}$ and results in $\rho(s_1,s_2)$, i.e., $\sigma = (\sigma_1, \ldots, \sigma_m) \in \Omega(\rho(s_1,s_2),\{s_1, s_2\})$, we can find a permutation $\xi=(\xi_1,\ldots,\xi_m)$ of nodes in $\rho(u_1,u_m)$ such that $\xi_i=\sigma_{m-i+1}$ for $i=1,\ldots,m$. In other words, $\xi$ can be interpreted as the \emph{reverse} infection sequence corresponding to $\sigma$. Then $\Gamma(u_1,u_m)$ is the set of all such reverse infection sequences corresponding to $\Omega(\rho(s_1,s_2),\{s_1, s_2\})$. We show an illustration of these definitions in Figure \ref{fig:sample_network}. Each term $|T_u(s)|$ in the product in the R.H.S.\ of \eqref{eqn:C_onesource} can be interpreted as the number of nodes that can be infected via $u$ once $u$ has been infected. Similarly, the sum in \eqref{eqn:q} is over all possible reverse infection sequences $\xi$ of the nodes in $\rho(u_1,u_m)$,  and each term $I_i(\xi;s_1,s_2)$ in the product within the sum is the number of nodes that can be infected once $\xi_{i+1},\ldots,\xi_m$ have been infected.

By utilizing Lemma \ref{lemma:C_twosources}, we can compute $\C{u,v}{G_n}$ for any two nodes $u$ and $v$ in $G_n$ by evaluating $|T_w(u,v)|$ for all nodes $w \in G_n$, and the quantity $q(u_1,u_m; u, v)$, where $\rho(u,v) = (u,u_1,\ldots,u_m,v)$. With Assumption \ref{assumption:boundeddeg_tree}, Algorithm \ref{algo:T} allows us to compute $f_w(u) = |T_w(u)|$ and $g_w(u)=\prod_{v \in T_w(u)} |T_v(u)|$ for all neighbors $u$ of $w$, and for all $w\in G_n$ in $O(n)$ time complexity. To do this, we first choose any node $r \in G_n$, and consider $G_n$ as a directed tree with $r$ as the root node, and with edges in $G_n$ pointing away from $r$. Let $\parent{w}$ and $\children{w}$ be the parent and the set of children of $w$ in the directed tree $G_n$, respectively. Starting from the leaf nodes, let each non-root node $w\in G_n$ pass two messages containing $f_w(\parent{w})$ and $g_w(\parent{w})$ to its parent. Each node stores the values of these two messages from each of its children, and computes its two messages to be passed to its parent. When $r$ has received all messages from its children, a reverse sweep down the tree is done so that at the end of the algorithm, every node $w \in G_n$ has stored the values $\{f_u(w), g_u(w) : u \in \mathcal{N}_{G_n}(w)\}$. The algorithm is formally described in Algorithm \ref{algo:T}. The last product term on the R.H.S.\ of \eqref{eqn:C_two} can then be computed using
\begin{align}\label{eqn:g}
g(s_1,s_2) = \prod_{w\in\rho(s_1,s_2)}\prod_{x\in \mathcal{N}_{G_n}(w)\backslash\rho(s_1,s_2)} g_x(w),
\end{align}
and taking its reciprocal.

\begin{algorithm}[!hbt]
\caption{Tree Sizes and Products Computation}
\label{algo:T}
\begin{algorithmic}[1]
\STATE{\textbf{Inputs}: $G_n$}
\STATE{Choose any node $r \in G_n$ as the root node.}
\FOR{$w\in G_n$}
\STATE{Store received messages $f_x(w)$ and $g_x(w)$, for each $x \in \children{w}$.}
\IF{$w$ is a leaf}
    \STATE{$f_w(\parent{w}) = 1$}
    \STATE{$g_w(\parent{w}) = 1$}
\ELSE
    \STATE{$f_w(\parent{w}) =\sum_{x \in \children{w}} f_x(w) + 1$}
    \STATE{$g_w(\parent{w}) =f_w(\parent{w}) \cdot \prod_{x \in \children{w}}g_x(w)$}
\ENDIF
\STATE{Store $f_{\parent{w}}(w) = n - f_w(\parent{w})$.}
\STATE{Pass $f_w(\parent{w})$ and $g_w(\parent{w})$ to $\parent{w}$.}
\ENDFOR

\STATE{Set $g_{\parent{r}}(r) = 1$.}
\FOR{$w\in G_n$}
\STATE{Store received message $g_{\parent{w}}(w)$ from $\parent{w}$.}
\IF{$w$ is not a leaf}
	\FOR{$x\in \children{w}$}
    \STATE{$g_w(x) =f_w(x)\cdot g_{\parent{w}}(w) \cdot \prod_{y \in \children{w} \backslash \{x\}} g_y(w)$}
    \STATE{Pass $g_w(x)$ to $x$.}
	\ENDFOR
\ENDIF
\ENDFOR
\end{algorithmic}
\end{algorithm}

To compute $\C{s_1,s_2}{G_n}$ in \eqref{eqn:C_two}, we still need to compute $q(u_1,u_m;s_1,s_2)$. The recurrence \eqref{eqn:q_recurs} allows us to compute $q(u_1,u_m;s_1,s_2)$ for all $s_1,s_2\in G_n$ in $O(n^2d_{*}^2)$ complexity, where $d_*$ is the maximum node degree. The computation proceeds by first considering each pair of neighbors $(u,v)$. Both nodes have at most $d_{*}$ neighbors each, so that we need to evaluate $q(u,v;s_1,s_2)$ for all $s_1\in \mathcal{N}_{G_n}(u)\backslash\rho(u,v)$ and $s_2\in\ \mathcal{N}_{G_n}(v)\backslash\rho(u,v)$. This requires $O(d_{*}^2)$ computations. The computed values and $T_{\rho(u,v)}(s_1,s_2)$ are stored in a hash table. In the next step, we repeat the same procedure for node pairs that are two hops apart, and so on until we have considered every pair of nodes in $G_n$. Note that for a path $(u_1,\ldots,u_m)$ and $s_1,s_2$ neighbors of $u_1$ and $u_m$ respectively, $q(u_1,u_m;s_1,s_2)$ can be computed in constant time from \eqref{eqn:q_recurs} as $q(u_2,u_m;s_1,s_2) = q(u_2,u_m;u_1,s_2)$ and $q(u_1,u_{m-1};s_1,s_2) = q(u_1,u_{m-1};s_1,u_m)$. A similar remark applies for the computation of $|T_{\rho(u_1,u_m)}(s_1,s_2)|$. In addition, each lookup of the hash table takes $O(1)$ complexity since $G_n$ is known and collision-free hashing can be used. Therefore, the overall complexity is $O(n^2d_{*}^2)$.
The algorithm to compute the infection sources estimator is formally given in Algorithm \ref{algo:TSE}. We call this the Two Source Estimation (TSE) algorithm, and it forms the basis of our algorithm for multiple sources estimation in the sequel.

\begin{algorithm}[!t]
\caption{Two Source Estimation (TSE)}
\label{algo:TSE}
\begin{algorithmic}[1]
\STATE{\textbf{Input}: $G_n$}
\STATE{Let $(s^*_1,s^*_2)$ be the maximizer of $\C{\cdot,\cdot}{G_n}$. Set $C^* = 0$.}
\FOR{$d = 1$ to diameter of $G_n$}
	\FOR{each $s_1 \in G_n$}
		\FOR{each $s_2$ such that $d(s_1,s_2) = d$}
			\STATE{Let $\rho(s_1,s_2) = (s_1,u_1,\ldots,u_{d-1},s_2)$.}
			\IF{$d = 1$}
				\STATE{$q(u_1,u_{d-1};s_1,s_2) = 1$.}	
			\ELSIF{$d = 2$}
				\STATE{Store $q(u_1,u_1;s_1,s_2) = |T_{u_1}(s_1,s_2)|^{-1}$ and $|T_{u_1}(s_1,s_2)|$.}
			\ELSE
				\STATE{Look up $|T_{\rho(u_1,u_{d-2})}(s_1,u_{d-1})|$, $q(u_2,u_{d-1};u_1,s_2)$, and $q(u_1,u_{d-2};s_1,u_{d-1})$.}
				\STATE{Store
				\begin{align*}
				|T_{\rho(u_1,u_{d-1})}(s_1,s_2)| = |T_{\rho(u_1,u_{d-2})}(s_1,u_{d-1})|\cdot |T_{u_{d-1}}(s_1,s_2)|.
				\end{align*}
				}
				\STATE{Store
				\begin{align*}
				q(u_1,u_{d-1};s_1,s_2)
				= \frac{q(u_2,u_{d-1};u_1,s_2) + q(u_1,u_{d-2};s_1,u_{d-1})}{|T_{\rho(u_1,u_{d-1})}(s_1,s_2)|}.
				\end{align*}
				}
			\ENDIF
			\STATE{Compute $g(s_1,s_2)$ from \eqref{eqn:g}.}
			\STATE{$\C{s_1,s_2}{G_n} = (n-2)! q(u_1,u_{d-1};s_1,s_2) / g(s_1,s_2)$.}
			\STATE{Update $(s^*_1,s^*_2)$ and $C^*$ if $\C{s_1,s_2}{G_n} > C^*$.}
		\ENDFOR
	\ENDFOR
\ENDFOR
\end{algorithmic}
\end{algorithm}

\subsection{Geometric Trees with Two Sources}\label{subsection:GeometricTrees}
In this section, we study the special case of geometric trees, propose an approximate estimator for geometric trees, and provide theoretical analysis for its performance. First, we give the definition of geometric trees and prove some of its key properties. Then, we derive a lower bound for $\C{S}{G_n}$, and propose an estimator based on this lower bound. We show that our proposed estimator is asymptotically correct, i.e., it identifies the actual infection sources with probability (conditioned on the infection sources) going to one as the infection graph $G_n$ becomes large. For mathematical convenience, instead of letting the number of infected nodes $n$ grow large, we let the time $t$ from the start of the infection process to our observation time become large.

The geometric tree network is defined in \cite{Shah2011} w.r.t.\ a single infection source. In the following, we extend this definition to the case where there are two sources. Let $S^* = \{s_1,s_2\}$ be the infection sources, and let $T'_u(s_1,s_2)$ be defined in the graph $G$ in the same way as $T_u(s_1,s_2)$ is defined for $G_n$. Let $\mathcal{N}_{G}(\rho(s_1,s_2))$ be the set of nodes that have a neighboring node in $\rho(s_1,s_2)$. For each node $u$, let $n(u,r)$ be the number of nodes in $T'_u(s_1,s_2)$ that are at a distance $r$ from $u$. We say that $G$ is a geometric tree if for all $u\in\mathcal{N}(\rho(s_1,s_2))$, we have
\begin{align}\label{equ:regularity_condition}
br^\alpha \leq n(u,r) \leq cr^\alpha,
\end{align}
where $\alpha,b$, and $c$ are fixed positive constants with $b \leq c$. The condition \eqref{equ:regularity_condition} implies that all trees defined w.r.t.\ the infection sources are growing polynomially fast at about the same rate. As we have assumed that the infection rates are homogeneous for every node, the resulting infection graph $G_n$ will also be approximately regular with high probability. We have the following properties for a geometric tree, whose proofs are in Appendix \ref{appendix:lemma:N}.

\begin{lemma}\label{lemma:N}
Suppose that $G$ is a geometric tree with two infection sources $S^*=\{s_1,s_2\}$. Let $\alpha, b$ and $c$ be fixed positive constants satisfying \eqref{equ:regularity_condition} for the geometric tree $G$. Let $t$ be the time from the start of the infection process to our observation time. For any $\epsilon \in (0,1)$, let $\mathcal{E}_t$ be the event that all nodes within distance $t(1-t^{-1/2+\epsilon})$ of either source nodes are infected, and no nodes greater than distance $t(1+t^{-1/2+\epsilon})$ of either source nodes are infected. Then, there exists $t_0$ such that for all $t \geq t_0$, $\P(\mathcal{E}_t) \geq 1 - \epsilon$. Furthermore, conditioned on $\mathcal{E}_t$, we have for all $u \in \mathcal{N}_G(s_1)\cup\mathcal{N}_G(s_2)$ or $u = \rho(s_1,s_2)\backslash S^*$,
\begin{align}\label{ineq:TN}
N_{\min}(t) \leq |T_u(s_1,s_2)| \leq N_{\max}(t),
\end{align}
where
\begin{align}
N_{\min}(t) & = \frac{b}{1+\alpha}\left(t-t^{\frac{1}{2}+\epsilon}-d(s_1,s_2)-2\right)^{\alpha+1}, \label{equ:N_min}
\end{align}
and
\begin{align}
N_{\max}(t) = \frac{c}{1+\alpha}\left(t+t^{\frac{1}{2}+\epsilon}\right)^{\alpha+1}. \label{equ:N_max}
\end{align}
In addition, for $t \geq t_0$, we have
\begin{align*}
\frac{N_{\min}(t)}{N_{\max}(t)} \geq \frac{b}{c}(1 - \epsilon).
\end{align*}
\end{lemma}

\begin{figure}[!tb] 
  \centering
  \psfrag{a}[][][0.8][0]{$s_1$}
  \psfrag{b}[][][0.8][0]{$s_2$}
  \psfrag{c}[][][0.8][0]{$u_1$}
  \psfrag{d}[][][0.8][0]{$u_m$}
  \psfrag{e}[][][0.8][0]{$x_1$}
  \psfrag{f}[][][0.8][0]{$x_2$}
  \includegraphics[width=0.5\textwidth]{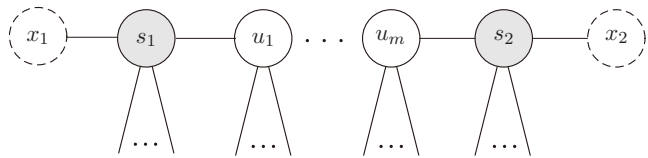}
  \caption{Addition of virtual nodes $x_1$ and $x_2$.}
  \label{fig:virtual_nodes}
\end{figure}

The infection sequences count in \eqref{eqn:C_two} is not amendable to analysis. In the following, we seek an approximation to simplify our analysis. For $s_1,s_2\in G_n$, suppose that $\rho(s_1,s_2) = (s_1,u_1,\ldots,u_m,s_2)$, with $p=|\rho(s_1,s_2)|=m+2$. Instead of computing $\C{s_1,s_2}{G_n}$, we consider a new infection graph $G'_n$ with two ``virtual'' nodes $x_i$, $i=1,2$ added, where $x_i$ is attached to $s_i$ (see Figure \ref{fig:virtual_nodes}). We now consider the infection sequence count $\C{x_1,x_2}{G'_n} \geq \C{s_1,s_2}{G_n}$. Since the trees rooted at $x_i$ are single node trees, we have
\begin{align*}
\C{x_1,x_2}{G'_n}
&= \C{s_1,x_2}{G'_n} + \C{x_1,s_2}{G'_n} \\
&\leq 2(n-1)\C{s_1,s_2}{G_n},
\end{align*}
where the last inequality follows because if $s_1$ and $x_2$ are sources, then $s_2$ can be inserted in any of at most $n-1$ positions in an infection sequence from $\Omega(G_n, \{s_1,s_2\})$, so that $\C{s_1,x_2}{G'_n} \leq (n-1)\C{s_1,s_2}{G_n}$. A similar argument holds for $\C{x_1,s_2}{G'_n} \leq (n-1)\C{s_1,s_2}{G_n}$.

Let $\xi^*=(\xi^*_1,\ldots,\xi^*_p)$ be a permutation of the nodes in $\rho(s_1,s_2)$ such that $|T_{\xi^*_i}(s_1,s_2)| \geq |T_{\xi^*_j}(s_1,s_2)|$ for all $1\leq i\leq j \leq p$, i.e., the nodes in $\xi^*$ are arranged in descending order of the size of the sub-trees rooted at them.  Let $I^*_i(s_1,s_2) = I_i(\xi^*;s_1,s_2)$ (cf.\ the definition in \eqref{eqn:I}) be the total number of nodes in the $i$ biggest sub-trees in $\{T_{u}(s_1,s_2) : u \in \rho(s_1,s_2)\}$. From Lemma \ref{lemma:C_twosources}, we have
\begin{align}
\C{x_1,x_2}{G'_n}
& \geq n! \cdot 2^{p-1} \prod_{i=1}^p I^*_i(s_1,s_2)^{-1} \prod_{u\in G_n\backslash \rho(s_1,s_2)} |T_u(s_1,s_2)|^{-1}, \label{ineq:C_virtual}
\end{align}
where the inequality holds because $|\Gamma(s_1,s_2)| = 2^{p-1}$, and each term in the sum on the R.H.S.\ of \eqref{eqn:q} is lower bounded by $\prod_{i=1}^p I^*_i(s_1,s_2)^{-1}$. We use the lower bound in \eqref{ineq:C_virtual} as a proxy for $\C{s_1,s_2}{G_n}$. However, we have used a very loose lower bound in \eqref{ineq:C_virtual}, so we propose the estimator
\begin{align}\label{estimator_twosource}
\tilde{S} = \arg\max_{s_1,s_2\in G_n} \hC{s_1,s_2}{G_n},
\end{align}
where
\begin{align}\label{equ:approximation_permulation_count}
\hC{s_1,s_2}{G_n} = n! \cdot Q(s_1,s_2)\prod_{u\in G_n\backslash\rho(s_1,s_2)} |T_u(s_1,s_2)|^{-1},
\end{align}
\begin{align*}
Q(s_1,s_2)=[2(1+\delta )]^{p-1} \prod_{i=1}^{p}I^*_i(s_1,s_2)^{-1},
\end{align*}
and $\delta$ is a fixed positive constant, to be chosen based on prior knowledge about the graph $G$. Algorithm \ref{algo:TSE} can be modified to find the maximizer for $\hC{\cdot,\cdot}{G_n}$. We call this the geometric tree TSE algorithm. The following result provides a way to choose $\delta$, and shows that our proposed estimator $\tilde{S}$ is asymptotically correct in a geometric tree. A proof is provided in Appendix \ref{appendix:theorem:TSE_detection_prob}.

\begin{theorem}\label{theorem:TSE_detection_prob}
Suppose that $G$ is a geometric tree with two infection sources $S^*=\{s_1^*,s_2^*\}$. Let $d_{\min}$ and $d_{\max}$ be constants such that $\deg_G(s_i) \in [d_{\min},d_{\max}]$ for $i=1,2$. Let $b$ and $c$ be fixed positive constants satisfying \eqref{equ:regularity_condition} for the geometric tree $G$. Suppose that
\begin{align}\label{dmin}
d_{\min} \geq \frac{3}{2} + \frac{c}{b}\sqrt{2d_{\max}}.
\end{align}
Then, for any $\delta$ in the non-empty interval
\begin{align}\label{deltacond}
\left(\frac{cd_{\max}}{b(d_{\min}-1)}-1,\frac{b(d_{\min}-2)}{2c}-1\right),
\end{align}
we have
\begin{align*}
\lim_{t\to\infty} \P(\tilde{S} = S^* \mid S^*) = 1.
\end{align*}
\end{theorem}
Theorem \ref{theorem:TSE_detection_prob} implies that if we know the constants governing the regularity condition \eqref{equ:regularity_condition} for $G$, we can choose a $\delta$ so that our estimator $\tilde{S}$ gives the true infection sources with high probability if the infection graph $G_n$ is large. The class of geometric trees as defined by \eqref{equ:regularity_condition} can be used to model various scenarios in practice, e.g., a tree spanning a wireless sensor network with nodes randomly scattered. However, the assumption \eqref{equ:regularity_condition} may also be overly strong for other applications. In Section \ref{sec:simulation_results}, we perform numerical studies to gain insights into the performance of our proposed estimator for different classes of tree networks.

\subsection{Unknown Number of Infection Sources}\label{subsec:UnknownSourceNumber}

In most practical applications, the number of infection sources is not known a priori. However, typically we may be able to guess the maximum number of infection sources $\kmax$, or we can choose a reasonable value of $\kmax$ depending on the size of the infection graph $G_n$. In this section, we present a \textit{heuristic} algorithm that allows us to estimate the infection sources with a given $\kmax$.

We first consider the instructive case where $\kmax=2$ and $G$ is a geometric tree. In this case, the number of infection sources can be either one or two. Suppose we run the geometric tree TSE algorithm on $G_n$. We have the following result, whose proof is in Appendix \ref{appendix:theorem:TSE_differentiate}.

\begin{theorem}\label{theorem:TSE_differentiate}
Suppose that there is a single infection source $s$ and $G$ is a geometric tree with \eqref{equ:regularity_condition} holding for all nodes $u$ that are neighbors of $s$. Suppose that $s$ has degree $\deg_G(s) \in [d_{\min}, d_{\max}]$, where $d_{\min}$ and $d_{\max}$ are positive constants satisfying \eqref{dmin}. Then, for any $\delta$ in the interval \eqref{deltacond}, the geometric tree TSE algorithm estimates as sources $s$ and one of its neighbors with probability (conditioned on $s$ being the infection source) going to $1$ as $t\to \infty$.
\end{theorem}

Theorem \ref{theorem:TSE_differentiate} implies that when there exists only one source, the geometric tree TSE algorithm finds two neighboring nodes, one of which is the true source. From Theorem \ref{theorem:TSE_detection_prob} and Assumption \ref{assumption:source_separation}, if there are two sources, our algorithm identifies the two source nodes, which are at least two hops from each other, with high probability. Therefore, by checking the distance between the two nodes identified by the geometric tree TSE algorithm, we can estimate the number of source nodes in the infection graph. This observation together with Theorem \ref{theorem:A} suggest the following heuristic.
\begin{enumerate}[(i)]
\item\label{it:IP} Randomly choose $\kmax$ nodes satisfying Assumption \ref{assumption:source_separation} as the infection sources and find a Voronoi partition for $G_n$. Use the SSE algorithm to find a source node for each infection region. Repeat these steps until for every region, the distance between estimated source nodes between iterations is below a fixed threshold or a maximum number of iterations is reached. We call this the Infection Partition (IP) Algorithm (see Algorithm \ref{algo:Infection_Region_Partition_Algorithm}).
\item For any two regions in the partition obtained from step \eqref{it:IP} that are connected by an edge in $G_n$, run the TSE algorithm in the combined region to find two source estimates. If the two estimates have distance less than $\tau$, we decrement the number of source nodes, and repeat step \eqref{it:IP}.
\item The above two steps are repeated until no two pairs of regions in the Voronoi partition can be combined. The formal algorithm is given as the Multiple Sources Estimation and Partitioning (MSEP) algorithm in Algorithm \ref{algo:multiple_sources}.
\end{enumerate}

\begin{algorithm}[!t]
\caption{Infection Partitioning (IP)}
\label{algo:Infection_Region_Partition_Algorithm}
\begin{algorithmic}[1]
\STATE{\textbf{Inputs}: An infection source set $S\tc{0} = \{s\tc{0}_i : i=1,\ldots,m\}$ in $G_n$.}

\STATE{\textbf{Iterations}:}
\FOR{$l=1$ to MaxIter}
\STATE{Run the Voronoi partitioning algorithm with centers in $S\tc{l-1}$ to obtain the infection partition $\cA\tc{l} = \cup_{i=1}^m A\tc{l}_i$.}

\FOR{$i=1$ to $m$}
\STATE{Run SSE algorithm in $A\tc{l}_i$ to obtain
\begin{align*}
s\tc{l}_i = \arg\max_{s\in A\tc{l}_i} \C{s}{A\tc{l}_i}.
\end{align*}
}\label{IP:SSE}
\ENDFOR
\STATE{$S\tc{l} := \{s\tc{l}_i : i=1,\ldots,m\}$}
\IF{$\max_{1\leq i\leq m} d(s\tc{l}_i,s\tc{l-1}_i) \leq \eta$ for some fixed small positive $\eta$}
\STATE{break}
\ENDIF
\ENDFOR
\RETURN{$(S\tc{l},\cA\tc{l})$}
\end{algorithmic}
\end{algorithm}

\begin{algorithm}[!hbt]
\caption{Multiple Sources Estimation and Partitioning (MSEP)}
\label{algo:multiple_sources}
\begin{algorithmic}[1]
\STATE{\textbf{Inputs}: $G_n$ and $\kmax$.}
\STATE{\textbf{Initialization}:}
\STATE{$k := \kmax$ and choose $S := \{s_1,\ldots,s_k\}$ randomly in $G_n$.}
\STATE{\textbf{Iterations}:}
\WHILE{$k > 1$}
\STATE{$(S,\cA) =$ Algorithm IP($S$)}
\STATE{$S' := S$}
\FORALL{regions $A_i$ and $A_j$ in the partition $\cA$ such that there exists an edge $(u,v)$ in $G_n$ with $u\in A_i$ and $v \in A_j$}\label{MSEP:TSE_Loop}
\STATE{Set $(u,v) =$ Algorithm TSE($A_i\cup A_j$).}\label{MSEP:TSE}
\IF{$d(u,v) < \tau$}
\STATE{Merge $A_i$ and $A_j$, set $s_i = u$ and discard $s_j$}
\STATE{$k := k - 1$}
\STATE{break}
\ENDIF
\ENDFOR
\IF{$S = S'$}
\STATE{break}
\ENDIF
\ENDWHILE
\RETURN{$(S,\cA)$}
\end{algorithmic}
\end{algorithm}

To compute the complexity of the MSEP algorithm, we note that since the IP algorithm is based on the SSE algorithm, it has complexity $O(n)$. For each value of $k = 1,\ldots, \kmax$ in the MSEP algorithm, there are $O(k^2)$ pairs of neighboring regions in the infection partition. For each pair of region, the TSE algorithm makes $O(n^2)$ computations. Summing over all $k=1,\ldots,\kmax$, the time complexity of the MSEP algorithm can be shown to be $O(\kmax^3 n^2)$. On the other hand, to compute $\C{S}{G_n}$ for $|S^*| = \kmax$ would require $O(n^{\kmax})$ computations.

\section{Identifying Infection Sources and Regions for General Graphs}\label{sec:source_estimation_general}

In this section, we generalize the MSEP algorithm to identify multiple infection sources in general graphs $G$. In \cite{Shah2011}, the SSE algorithm is extended to general graphs when it is known that there is only a single infection source in the network using a heuristic. The algorithm first chooses a node $s$ of $G_n$ as the root node, and generates a spanning tree $T_{\mathrm{bfs}}(s,G_n)$ of $G_n$ rooted at $s$ using the breadth-first-search (BFS) procedure. The SSE algorithm is then applied on this spanning tree to compute $\C{s}{T_{\mathrm{bfs}}(s,G_n)}$. In addition, the infection sequences count is weighted by the likelihood of the BFS tree. This is repeated using every node in $G_n$ as the root node, and the node $\hat{s}$ with the maximum weighted infection sequences count is chosen as the source estimator, i.e.,
\begin{align*}
\hat{s} &= \arg \max_{v\in G_n} P(\sigma_v \mid v) \C{s}{T_{\mathrm{bfs}}(v,G_n)},
\end{align*}
where $\sigma_v$ is the sequence of nodes that corresponds to an infection spreading from $v$ along the BFS tree. It can be shown that this algorithm has complexity $O(n^2)$. For further details, the reader is referred to \cite{Shah2011}. We call this algorithm the SSE-BFS algorithm in this paper.

We adapt the MSEP algorithm for general graphs using the same BFS heuristic. Specifically, we replace the SSE algorithm in line \ref{IP:SSE} of the IP algorihm with the SSE-BFS algorithm. In addition, in line \ref{MSEP:TSE}, we run the TSE algorithm on $T_{\mathrm{bfs}}(s_i,A_i) \cup T_{\mathrm{bfs}}(s_j,A_j)$, where the two BFS trees are connected by randomly selecting an edge $(u,v)$ in $G_n$ with $u\in T_{\mathrm{bfs}}(s_i,A_i)$ and $v \in T_{\mathrm{bfs}}(s_j,A_j)$. We call this modified algorithm the MSEP-BFS algorithm. Since the worst case complexity for the SSE-BFS algorithm is $O(n^2)$, the complexity of the MSEP-BFS algorithm can be shown to be $O(\kmax^3 n^2)$, which is the same complexity as the MSEP algorithm. To verify the effectiveness of the MSEP-BFS algorithm, we conduct simulations on both synthetic and real world networks in Section \ref{sec:simulation_results}.

\section{Simulation Results and Tests}\label{sec:simulation_results}
In this section, we present results from simulations and tests on real data to verify our proposed algorithms. We first consider geometric tree networks and regular tree networks with various numbers of infection sources, and then we present results on small-world networks and a real world power grid network. We also apply our algorithms to the contact tracing data obtained during the SARS outbreak in Singapore in 2003 \cite{Goh2006} and the Arizona-Southern California cascading power outages in 2011\cite{Outage2012}.

\subsection{Synthetic Networks}\label{subsec:simulation_synthetic}

\begin{figure}[!t] 
  \centering
  \psfrag{a}[][][0.9][0]{$|S^*|$}
  \includegraphics[width=0.45\textwidth]{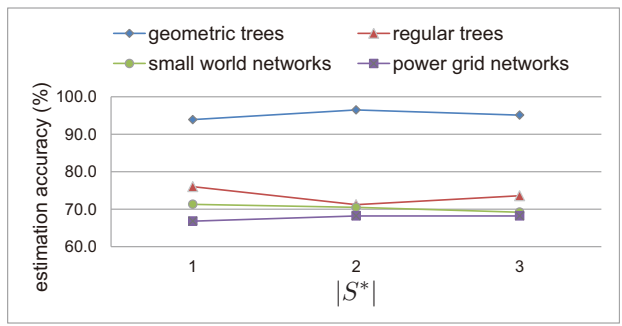}
  \caption{Estimating the number of infection source nodes.}
  \label{fig:estimated_number_of_source}
\end{figure}

We perform simulations on geometric trees, regular trees, and small-world networks. The number of infection sources $|S^*|$ are chosen to be 1, 2, or 3, and we set $\kmax=3$. For each type of network and each number of infection sources, we perform $1000$ simulation runs with 500 infected nodes. We randomly choose infection sources satisfying Assumption \ref{assumption:source_separation} and obtain the infection graph by simulating the infection spreading process using the SIR model. Finally, the MSEP or MSEP-BFS algorithm for tree networks and small-world networks respectively, is applied to the infection graph to estimate the number and locations of the infection sources. The estimation results for the number of infection sources $|\hat{S}|$ in different scenarios are shown in Figure \ref{fig:estimated_number_of_source}. It can be seen that our algorithm correctly finds the number of infection sources more than $93\%$ of the time for geometric trees, and more than $71\%$ of the time for regular trees. The accuracy of about 69.2\% for small-world networks is worse than that for the tree networks, as the infection tree for a small-world network has to be estimated using the BFS heuristics, thus additional errors are introduced into the procedure.

\begin{table*}[!t]
  \begin{center}
    \caption{Performance comparions.}  \label{table:average_error_distances}
    \begin{tabular}{c|c||c|c|c|c|c|c|c|c|c|}
    \cline{1-9}
    \multicolumn{2}{|c||}{Simulation settings} & Average & \multicolumn{5}{c|}{Average error distance $\Delta$} & Average minimum \\ \cline{1-2} \cline{4-8}
    \multicolumn{1}{|c|}{\multirow{2}{*}{network topology}} & \multicolumn{1}{c||}{\multirow{2}{*}{$|S^*|$}} & \multicolumn{1}{c|}{\multirow{1}{*}{diameter}} & \multicolumn{2}{c|}{\multirow{1}{*}{MSEP/MSEP-BFS}} & \multicolumn{3}{c|}{\multirow{1}{*}{nSSE}} & \multicolumn{1}{c|}{\multirow{1}{*}{infection region}} \\ \cline{4-8}
    \multicolumn{1}{|c|}{} & \multicolumn{1}{c||}{} & \multicolumn{1}{c|}{of $G_n$} & $\eta = 0$ & \multicolumn{1}{c|}{\multirow{1}{*}{$\eta = \textrm{diameter}$}} & $\eta = 0$ & \multicolumn{1}{c|}{\multirow{1}{*}{$\eta = \textrm{diameter}$}} & known $|S^*|$ & \multicolumn{1}{c|}{covering percentage (\%)} \\ \cline{1-9}
    \multicolumn{1}{|c|}{\multirow{2}{*}{geometric trees}} & 2 & 63.7 & 0.61 & 1.72 & 9.65 & 30.16 & 12.85 & 97.06\\ \cline{2-9}
    \multicolumn{1}{|c|}{} & 3 & 66.2 & 0.91 & 2.42 & 7.69 & 29.95 & 14.84 & 89.77   \\ \cline{1-9}
    \multicolumn{1}{|c|}{\multirow{2}{*}{regular trees}} & 2 & 40.5 & 0.84 & 6.07 & 4.50 & 17.70 & 6.13  & 73.82    \\ \cline{2-9}
    \multicolumn{1}{|c|}{} & 3 & 43.7 & 0.94 & 6.24 & 3.39 & 17.47 & 6.59  & 65.95    \\ \cline{1-9}
    \multicolumn{1}{|c|}{\multirow{2}{*}{small-world networks}} & 2 & 35.5 & 2.95 & 8.19 & 5.40 & 17.13 & 8.28   & 76.62    \\ \cline{2-9}
    \multicolumn{1}{|c|}{} & 3 & 40.9 & 2.58 & 8.18 & 4.99 & 18.56 & 10.37  & 60.69    \\ \cline{1-9}
    \multicolumn{1}{|c|}{\multirow{2}{*}{power grid network}} & 2 & 27.3 & 3.65 & 7.39 & 5.50 & 14.66 & 7.89  & 70.29   \\ \cline{2-9}
    \multicolumn{1}{|c|}{} & 3 & 30.8 & 2.85 & 8.47 & 4.71 & 14.75 & 8.89  & 59.95    \\ \cline{1-9}
    \end{tabular}
	\end{center}
\end{table*}

When there are more than one infection sources, we compare the performance of the MSEP algorithm with a naive estimator based on the SSE algorithm. We call this the nSSE algorithm. Specifically, in the estimator for tree networks, we first compute $\C{u}{G_n}$ for all nodes $u\in G_n$, and choose the $|S^*|$ nodes with the largest counts as the source nodes. In non-tree networks, we use the SSE-BFS algorithm. Since the nSSE algorithm can not estimate $|S^*|$, we consider two variants. In the first variant, we assume the nSSE algorithm has prior knowledge of $|S^*|$. In the second variant, we guess $|S^*|$ by choosing uniformly from $\{1,\ldots,\kmax\}$.

\begin{figure*}[!ht]
  \centering

  \psfrag{a}[l][][0.6][0]{$|S^*|=2$, MSEP}
  \psfrag{b}[l][][0.6][0]{$|S^*|=2$, nSSE}
  \psfrag{c}[l][][0.6][0]{$|S^*|=3$, MSEP}
  \psfrag{d}[l][][0.6][0]{$|S^*|=3$, nSSE}
  \psfrag{e}[l][][0.6][0]{$|S^*|=2$, MSEP-BFS}
  \psfrag{g}[l][][0.6][0]{$|S^*|=2$, nSSE}
  \psfrag{h}[l][][0.6][0]{$|S^*|=3$, MSEP-BFS}
  \psfrag{k}[l][][0.6][0]{$|S^*|=3$, nSSE}

  \subfigure[Geometric trees.]{
    \label{fig:geometric_error_distance}
    \includegraphics[width=0.45\textwidth]{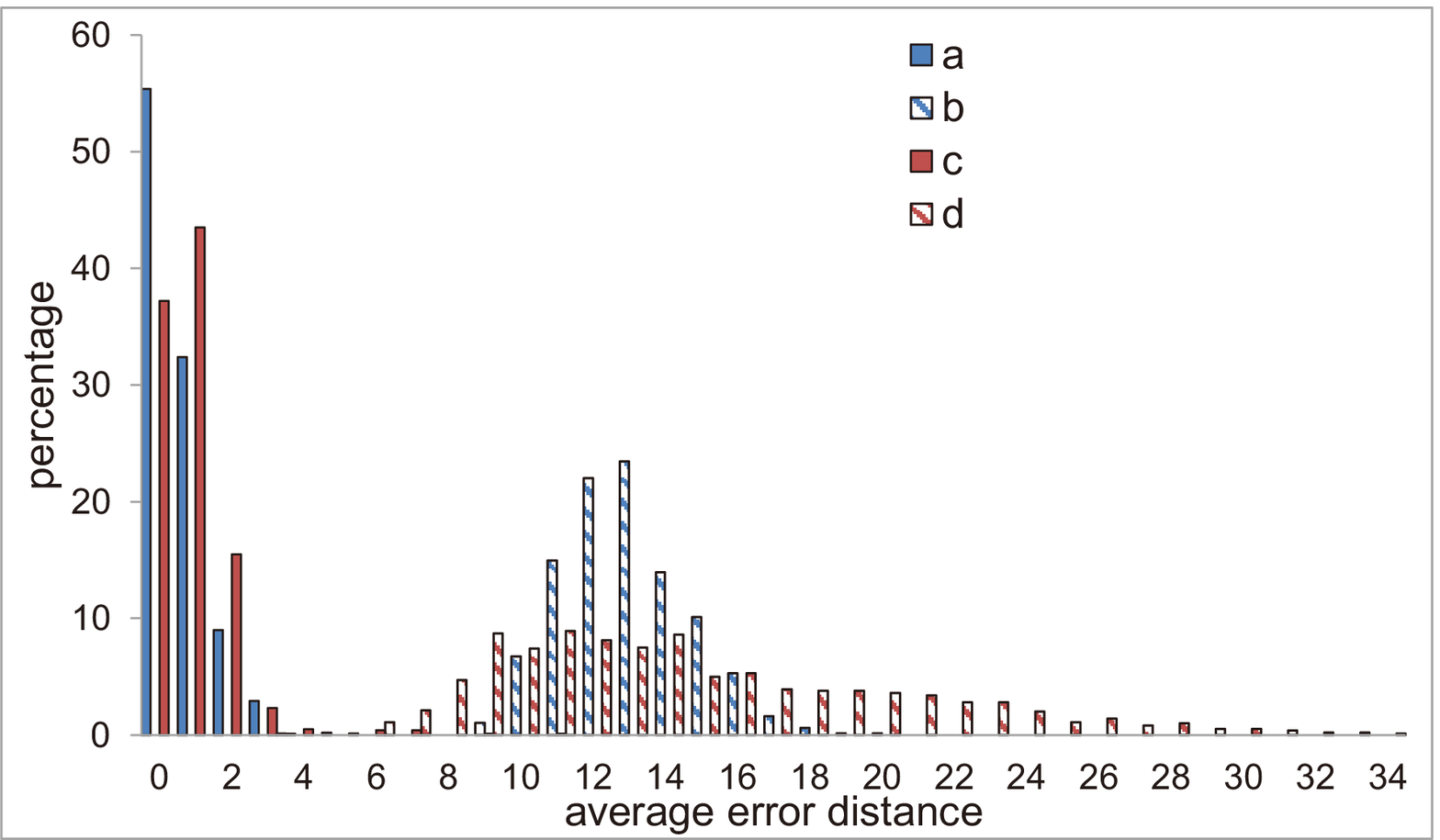}}
  \hspace{0.1in}
  \subfigure[Regular trees.]{
    \label{fig:regular_error_distance}
    \includegraphics[width=0.45\textwidth]{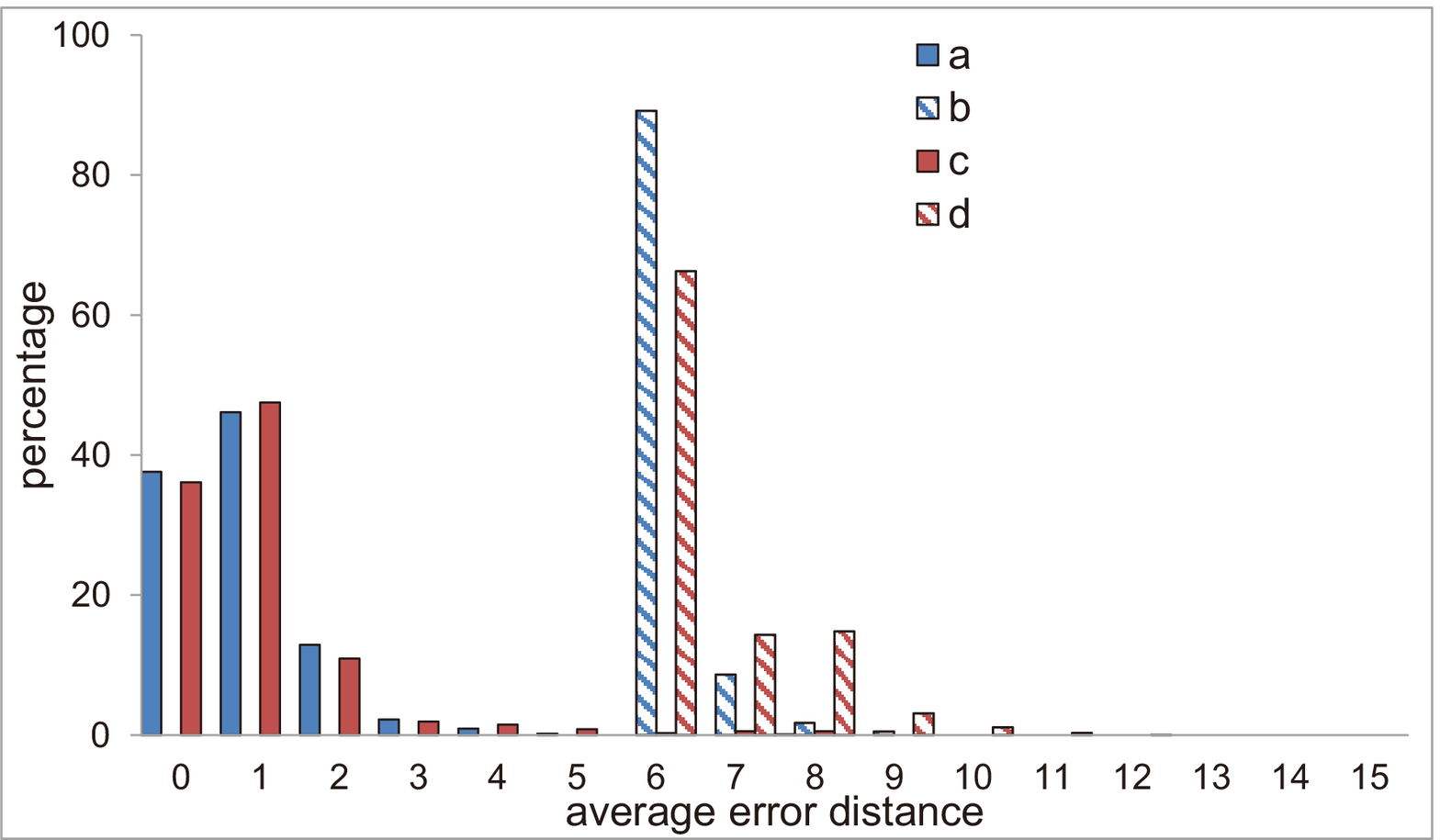}}
  \hspace{0.1in}
  \subfigure[Small-world networks.]{
    \label{fig:small_world_error_distance}
    \includegraphics[width=0.45\textwidth]{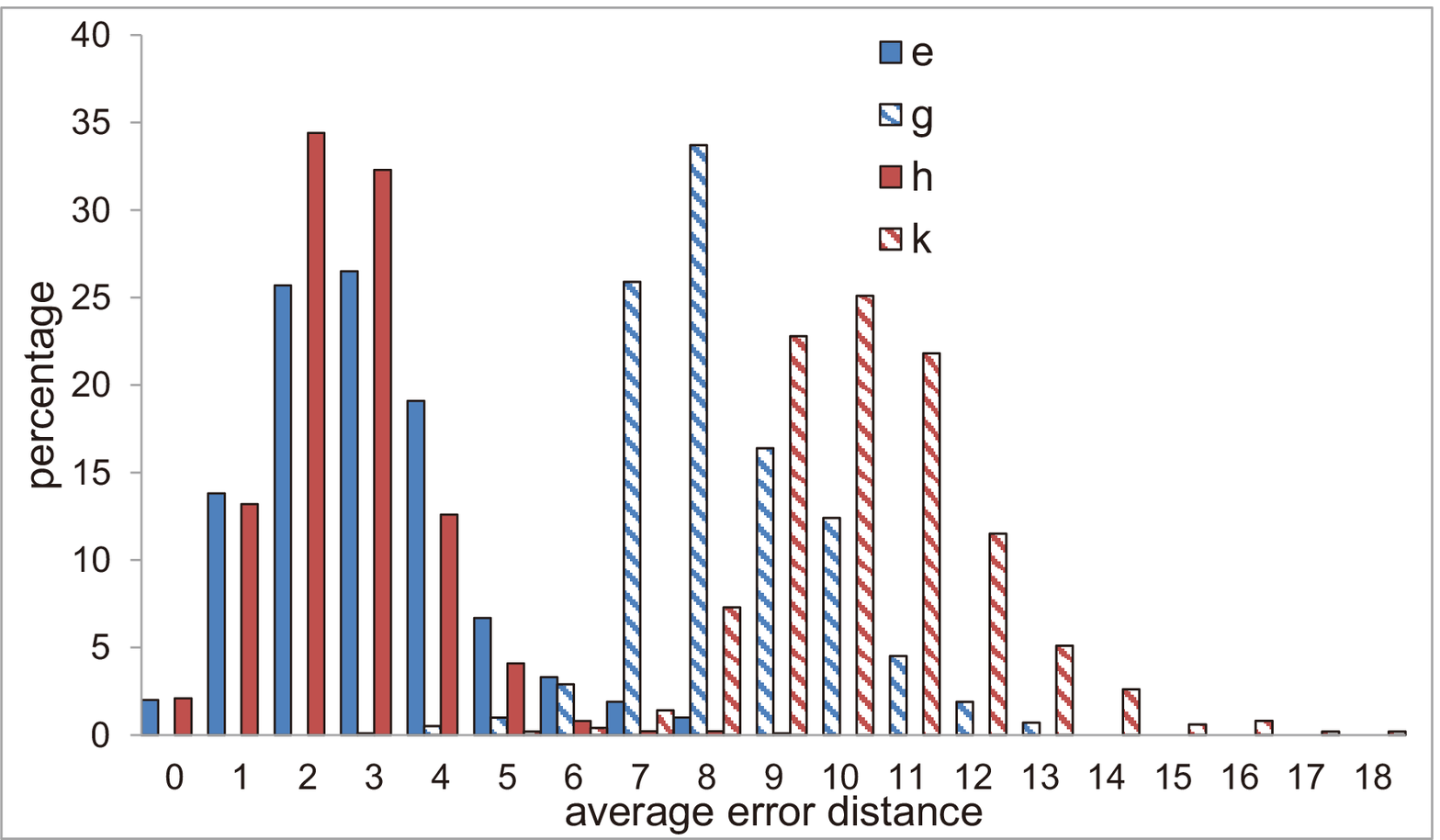}}
\hspace{0.1in}
  \subfigure[US power grid network.]{
    \label{fig:power_grid_error_distance}
    \includegraphics[width=0.45\textwidth]{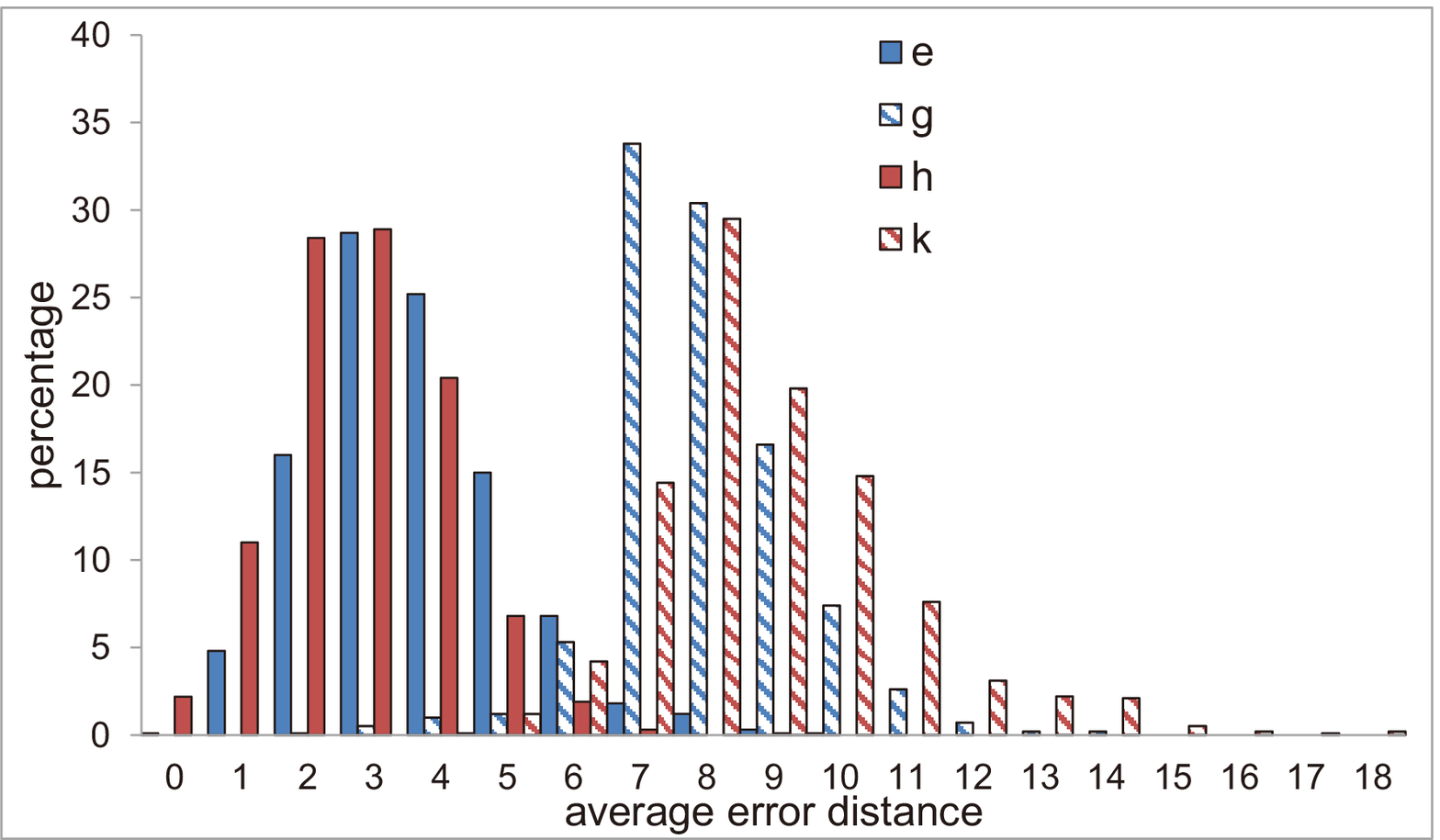}}
\caption{Histogram of the average error distances for various networks. We assume $\eta=0$ and that the nSSE algorithm has prior knowledge of the number of infection sources.}
\label{fig:error_distance}
\end{figure*}

To quantify the performance of each algorithm, we first match the estimated source nodes $\hat{S} = \{\hat{s}_i: i=1,\ldots,|\hat{S}|\}$ with the actual sources $S^*$ so that the sum of the error distances between each estimated source and its match is minimized. Let this matching be denoted by the function $\pi$, which matches each actual source $s_i$ to $\hat{s}_{\pi(i)}$. If we have incorrectly estimated the number of infection sources, i.e., $|\hat{S}|\neq |S^*|$, we add a penalty term to this sum. The average error distance is then given by
\begin{align*}
\Delta &= \frac{1}{|S^*|}\left(\sum_{i=1}^{\min(|S^*|,|\hat{S}|)} d(\hat{s}_{\pi(i)}, s_i) + \eta\left||\hat{S}|-|S^*| \right| \right),
\end{align*}
where $\eta$ is a penalty weight for incorrectly estimating the number of infection sources. For different applications, we may assign different values to $\eta$ depending on how important it is to estimate correctly the number of infection sources. In our simulations, we consider the cases where $\eta=0$, and where $\eta$ is the diameter of the infection graph. The average error distances for the different types of networks are provided in Table \ref{table:average_error_distances}. Clearly, the MSEP/MSEP-BFS algorithm outperforms the nSSE algorithm, even when the nSSE algorithm has prior knowledge of the number of sources. When $|S^*|$ is known a priori, the performance of the nSSE algorithm deteriorates with increasing $|S^*|$. This is to be expected as the SSE algorithm assumes that the node with the largest infection sequence count is the only source, and this node tends to be close to the distance center \cite{Sabidussi1966} of the infection graph. The histogram of the average error distances when $\eta=0$ are shown in Figure \ref{fig:error_distance}.

The MSEP/MSEP-BFS algorithm also estimates the infection region of each source. To evaluate its accuracy, we first perform the matching process described previously. Let the true infection region of $s_i$ be $A_{n,i}$ and the estimated infection region of $\hat{s}_{\pi(i)}$ be $\hat{A}_{n,i}$, where we set $\hat{A}_{n,i}=\emptyset$, if we have underestimated the number of sources and $s_i$ is unmatched. We define the correct infection region covering percentage for $s_i$ as the ratio between $|\hat{A}_{n,i} \cap A_{n,i}|$ and $|A_{n,i}|$, and we compute the minimum (or worst case) infection region covering percentage as
\begin{align*}
\operatornamewithlimits{\mathrm{min}}_{i \in \{1, \cdots, |S^*|\}} \frac{|\hat{A}_{n,i} \cap A_{n,i}|}{|A_{n,i}|}.
\end{align*}
This is then averaged over all simulation runs. We find that the average minimum infection region covering percentage is more than 59\% for all networks, as shown in Table \ref{table:average_error_distances}.

\subsection{Real World Networks}\label{subsec:simulation_real}
We verify the performance of the MSEP-BFS algorithm on the western states power grid network of the United States \cite{Watts1998}. We simulate the infection spreading process on the power grid network, which contains $4941$ nodes. For each simulation run, 1, 2 or 3 infection sources are randomly chosen from the power grid network under Assumption \ref{assumption:source_separation}, and the spreading process is simulated so that a total of $500$ nodes are infected. For each value of $|S^*|$, $1000$ simulation runs are performed. The simulation results are shown in Figures \ref{fig:estimated_number_of_source} and \ref{fig:error_distance}\subref{fig:power_grid_error_distance}, and Table \ref{table:average_error_distances}. We see that the MSEP-BFS algorithm outperforms the nSSE algorithm in every scenario. The average infection region covering percentage is above 59\%.

\subsection{Tests on Real Data}\label{subsec:experiments}

\begin{figure}[!htb] 
  \centering
  \includegraphics[width=0.28\textwidth]{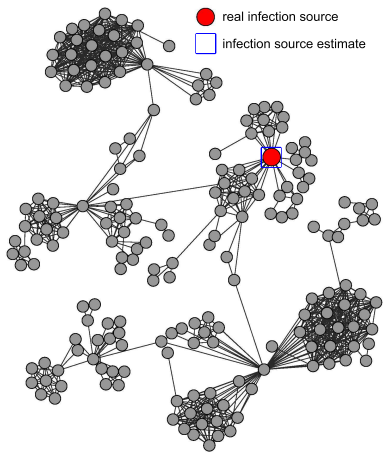}
  \caption{Illustration of a cluster of the SARS infection network with a single source.}
  \label{fig:sars}
\end{figure}

In order to get some insights in the performance of the MSEP-BFS algorithm in real infection spreads, we conduct two tests on real infection spreads data. We first apply the MSEP-BFS algorithm to to a network of nodes that represent the individuals who were infected with the SARS virus during an epidemic in Singapore in the year 2003. The data is collected using contact tracing of patients \cite{Goh2006}, where an edge between two nodes indicate that there is some form of interaction or relationship between the individuals (e.g., they are family members, classmates, colleagues, or commuters who shared the same public transport system). A part of the SARS infection network corresponding to a cluster of 193 patients is shown in Figure \ref{fig:sars}. We test the MSEP-BFS algorithm on the network in Figure \ref{fig:sars}, assuming that there are at most $\kmax = 3$ infection sources. It turns out that the MSEP-BFS algorithm correctly estimates the number of infection sources to be one, and correctly identifies the real infection source.

\begin{figure}[!htb] 
  \centering
  \includegraphics[width=0.3\textwidth]{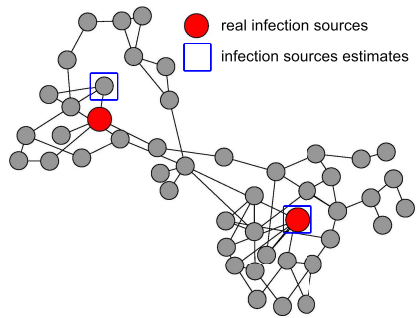}
  \caption{Illustration of the main affected power network with two infection sources.}
  \label{fig:power_outage}
\end{figure}

We next consider the Arizona-Southern California cascading power outages in 2011\cite{Outage2012}. The affected power network is represented by a graph where a node represents a key facility (substation or generating plant) affected by an outage, and an edge between two nodes indicate that there is a transmission line between these two facilities. The cascading outage starts with the loss of a single transmission line. However, as indicated in \cite{Outage2012}, this transmission line alone would not cause a cascading outage. After the loss of this transmission line, instantaneous power flow redistributions led to large voltage deviations, resulting in the nuclear units at San Onofre Nuclear Generating Station being taken off the power grid. The failures of these two key facilities together serve as the main causes of the subsequent cascading outages, so these two facilities are considered as the two infection sources. The main affected power network containing 48 facilities is shown in Figure \ref{fig:power_outage}. We test the MSEP-BFS algorithm on the network in Figure \ref{fig:power_outage}, and assume that there are at most $\kmax = 3$ infection sources. We can see that the MSEP-BFS algorithm correctly estimates the number of infection sources to be two. We also found one of the sources correctly, and one estimate 1 hop away from the real source.

\section{Conclusion}\label{sec:conclusion}

We have derived estimators for the infection sources and regions when the number of infection sources is bounded but unknown a priori. The estimators are based only on knowledge of the infected nodes and their underlying network connections. We provide an approximation for the infection source estimator for the class of geometric trees, and when there are at most two sources in the network. We show that this estimator asymptotically correctly identifies the infection sources when the number of infected nodes grows large. We also propose an algorithm that estimates the number of source nodes, and identify them and their respective infection regions for general infection graphs. Simulation results on geometric trees, regular trees, small-world networks, the US power grid network, and experimental results on the SARS infection network and cascading power outages show that our proposed estimation procedure performs well in general, with an average error distance of less than 4. The estimation accuracy of the number of source nodes is over 65\% in all the networks we consider, with the geometric tree networks having an accuracy of over 90\%. Furthermore, the minimum infection region covering percentage is more than 59\% for all networks. Our estimation procedure assumes only knowledge of the underlying network connections. In practical applications where more information about the infection process is available, a more accurate and intelligent guess of the number of infection sources can be made.

In this paper, we have adopted a simple SI infection model with homogeneous spreading rates, allowing us to derive analytical results that provide useful insights into infection source estimation for practical networks. However, this simplistic diffusion model does not adequately capture the real world dynamics of many networks. Future research includes the use of richer diffusion models that allow the inclusion of drifts and other dynamics in the infection spreading process, and tools from statistics to approximate optimal estimators for the infection sources. Our proposed algorithms find a set of nodes most likely to infect or influence a network, and are thus potentially useful for various practical applications. For example, our algorithm may be integrated with non-model-based consensus methods \cite{Hu2011,Hu2012} to design multi-agent control systems that uses only a small subset of agents as controllers. In cloud-centric media platforms \cite{Wen2011}, variants of our proposed algorithm may be used for intelligent content cache management. These are all areas of future research.

\appendices

\section{Proof of Theorem \ref{theorem:A}}\label{appendix:theorem:A}
Let nodes that are infected by source $s_i$ be colored with color $i$, with $i=1,\ldots,k$. Then a partition $\cA_n$ corresponds to a coloring of the graph $H_n$, and to quantify the probability of a partition, it is sufficient to consider only infection sequences in the graph $H_n$. We have
\begin{align}\label{eqn:cA}
P(\cA_n \mid S, G_n) = \sum_{\sigma\in\Omega(H_n,S,\cA_n)} P(\sigma \mid S),
\end{align}
where
\begin{align*}
\Omega(H_n,S,\cA_n) = \{\sigma\in\Omega(H_n,S) : \sigma\cap A_{n,i} \textrm{ is an infection sequence, for all $i=1,\ldots,k$}.\},
\end{align*}
and $\sigma\cap A_{n,i}$ is the subsequence of $\sigma$ containing only nodes that are in $A_{n,i}$.

Let $h = |H_n| - k$, and consider an infection sequence $\sigma = (\sigma_1,\ldots,\sigma_h) \in\Omega(H_n,S,\cA_n)$. Let the set of edges connecting susceptible nodes to infected nodes be called the susceptible edge set. We have assumed that the infection times of susceptible nodes are independent and identically exponentially distributed. Therefore, given the infection sequence $\sigma_1,\ldots,\sigma_{l-1}$, the next edge along which the infection is spread is chosen uniformly at random from the susceptible edge set at time index $l-1$. Since $H_n$ is a tree where all nodes except those in $S$ have degree 2, after infection of a new node, the susceptible edge set size remains the same except in the case where the infected node is the last node to be infected amongst those on a path connecting two infection sources. In that case, the susceptible edge set size reduces by 2. Let $J_\sigma$ be the set of indices of the last infected nodes on every path connecting infection sources. Letting $n_l = 1$ if $l \notin J_\sigma$ and 2 otherwise, we then have  \begin{align}
P(\sigma \mid S) &= \prod_{l=1}^{h} n_l p_l(\sigma \mid H_n, S) \nonumber\\
&= 2^p \prod_{l=1}^{h} p_l(\sigma \mid H_n, S) \label{eqn:A}
\end{align}
where $p$ is the number of paths connecting infection sources, and
\begin{align}\label{eqn:pl_tree}
p_l(\sigma \mid H_n, S) = \left(\sum_{s\in S} \deg_{H_n}(s) - 2 \sum_{j\in J_\sigma} \indicator{j < l} \right)^{-1}.
\end{align}

Choose two sources $s_i$ and $s_j$ and let $m$ be the number of nodes in the path $\rho(s_i,s_j)$ connecting $s_i$ and $s_j$, excluding the source nodes. Suppose that $r > \ceil{m/2}$ nodes in this path have color $i$. Construct a new coloring $\cA'_n$ so that $\ceil{m/2}$ nodes in $\rho(s_i,s_j)$ closest to $s_i$ have color $i$ and the rest have color $j$. The rest of the nodes in $\cA'_n$ have the same colors as that in $\cA_n$. Each infection sequence $\sigma\in\Omega(H_n,S,\cA_n)$ corresponds to an infection sequence $\sigma'\in\Omega(H_n,S,\cA'_n)$, where the last $x = r - \ceil{m/2}$ color-$i$ nodes in $\sigma$ become the last $x$ color-$j$ nodes in $\sigma'$. From \eqref{eqn:pl_tree}, we have $p_l(\sigma \mid H_n, S) = p_l(\sigma' \mid H_n, S)$ for all $l$. Since $\binom{m}{\ceil{m/2}} \geq \binom{m}{r}$, we have $|\Omega(H_n,S,\cA'_n)| \geq |\Omega(H_n,S,\cA_n)|$, therefore \eqref{eqn:cA} yields $P(\cA'_n \mid S,G_n) \geq P(\cA_n \mid S,G_n)$.

The same argument can be repeated a finite number of times for all paths in $H_n$ connecting infection sources. This shows that the estimator $\hat{\cA}_n(S)$ is a Voronoi partition of $G_n$, and the proof is complete.

\section{Proof of Lemma \ref{lemma:C_twosources}}\label{appendix:lemma:C_twosources}
To simplify notations, we write $T_u(s_1,s_2)$ as $T_u$, with the implicit understanding that all trees are defined w.r.t.\ $\{s_1,s_2\}$. The number of infection sequences can be found by counting the number of ways to form such a sequence. The $n-2$ slots in a sequence are occupied by nodes from $T_{s_i}\backslash\{s_i\}$, $i=1,2$, and $T_{\rho(u_1,u_m)}$. Therefore, we have
\begin{align*}
\C{s_1,s_2}{G_n}
&= (n-2)!\prod_{i=1}^2\frac{\C{s_i}{T_{s_i}}}{(|T_{s_i}|-1)!}\cdot
\frac{R(u_1,u_m)}{|T_{\rho(u_1,u_m)}|!} \\
&= \frac{(n-2)!}{|T_{\rho(u_1,u_m)}|!} \cdot R(u_1,u_m) \cdot \prod_{\substack{v\in T_{s_i},i=1,2 \\ v\ne s_1,s_2}} |T_v|^{-1},
\end{align*}
where $R(u_i,u_j)$ for $i \leq j$ is the number of ways of permuting the nodes in $T_{\rho(u_i,u_j)}$ such that the infection sequence property is maintained, and the last equality follows from Lemma \ref{lemma:C_onesource}. To simplify the notations, for $1\leq i\leq j \leq m$, let
\begin{align*}
J(u_i,u_j) = \prod_{v\in T_{\rho(u_i,u_j)}\backslash\rho(u_i,u_j)} |T_v|^{-1}.
\end{align*}
For example, from Lemma \ref{lemma:C_onesource}, we have $\C{u_i}{T_{u_i}} = (|T_{u_i}|-1)! J(u_i,u_i)$.
In the following, we show that for $1\leq i\leq j \leq m$,
\begin{align}\label{eqn:R}
R(u_i,u_j) = |T_{\rho(u_i,u_j)}|! \cdot q(u_i,u_j;s_1,s_2) \cdot J(u_i,u_j).
\end{align}
The proof proceeds by induction on $j-i$. If $j=i$, we have $R(u_i,u_i)= \C{u_i}{T_{u_i}}$ and the claim follows from Lemma \ref{lemma:C_onesource}. Suppose that the claim \eqref{eqn:R} holds for all nodes $u_k$ and $u_p$ such that $p-k < j-i$. The number of permutations $R(u_i,u_i)$ can be computed by considering a sequence with $m=|T_{\rho(u_i,u_j)}|$ slots. The first slot can be filled with either $u_i$ or $u_j$. Therefore, we have
\begin{align*}
R(u_i,u_j)
&= (m-1)!\left(\frac{\C{u_i}{T_{u_i}}}{(|T_{u_i}|-1)!} \frac{R(u_{i+1},u_j)}{|T_{\rho(u_{i+1},u_j)}|!} + \frac{\C{u_j}{T_{u_j}}}{(|T_{u_j}|-1)!} \frac{R(u_{i},u_{j-1})}{|T_{\rho(u_{i},u_{j-1})}|!} \right) \\
&= (m-1)! \left( J(u_i,u_i) q(u_{i+1},u_j;s_1,s_2) J(u_{i+1},u_j) + J(u_j,u_j) q(u_{i},u_{j-1};s_1,s_2) J(u_{i+1},u_j) \right) \\
&= (m-1)! \left(q(u_{i+1},u_j;s_1,s_2) + q(u_{i},u_{j-1};s_1,s_2)\right)J(u_i,u_j),
\end{align*}
where the penultimate equality follows from the inductive hypothesis and Lemma \ref{lemma:C_onesource}, and the last equality follows by noting that $J(u_i,u_i)J(u_{i+1},u_j)=J(u_j,u_j)J(u_{i+1},u_j)=J(u_i,u_j)$. The claim \eqref{eqn:R} now follows from \eqref{eqn:q_recurs}.
Finally, \eqref{eqn:q} follows by an inductive argument using \eqref{eqn:q_recurs}, which we omit. The proof is now complete.

\section{Proof of Lemma \ref{lemma:N}}\label{appendix:lemma:N}
The proof follows easily from Theorems 5 and 6 of \cite{Shah2011}. Consider the infection spreading along a path in $G_n$. Let $\Pi(t)$ be the counting process of the number of infected nodes in this path. The process $\Pi(t)$ consists of exponentially distributed arrivals with rate 1, and at most one arrival with rate 2 if the path is between the two infection sources. Let $\Pi_1(t)$ be a unit rate Poisson process corresponding to the rate 1 arrivals. Then $\Pi_1(t) \leq \Pi(t) \leq \Pi_1(t)+1$. From Theorem 6 of \cite{Shah2011}, we have for any positive $\gamma < 0.2$,
\begin{align*}
\P(\Pi(t) \leq t(1-\gamma)) &\leq \P(\Pi_1(t) \leq t(1-\gamma)-1) \leq \exp\left(-\ofrac{4}t(\gamma+\ofrac{t})^2\right), \\
\P(\Pi(t) \geq t(1+\gamma)) &\leq \P(\Pi_1(t) \geq t(1+\gamma)) \leq \exp\left(-\ofrac{4}t\gamma^2\right).
\end{align*}
The rest of the proof is the same as that of Theorem 5 of \cite{Shah2011}, and the proof is complete.

\section{Proof of Theorem \ref{theorem:TSE_detection_prob}}\label{appendix:theorem:TSE_detection_prob}

We first show that under \eqref{dmin}, the interval \eqref{deltacond} is non-empty. The condition \eqref{dmin} implies that
\begin{align*}
d_{\min} > \frac{3}{2} + \sqrt{2d_{\max} \frac{c^2}{b^2} - \ofrac{4}},
\end{align*}
which after some algebraic manipulations yields
\begin{align*}
& b^2(d_{\min}-1)(d_{\min}-2) > 2c^2d_{\max}, \\
& 1 \leq \frac{cd_{\max}}{b(d_{\min}-1)} < \frac{b(d_{\min}-2)}{2c}.
\end{align*}
Therefore \eqref{deltacond} is a non-empty interval. Fix a $\delta$ in the interval. Then for all $\epsilon > 0$ sufficiently small, we have
\begin{align*}
\frac{b(d_{\min}-1)(1+\delta)}{cd_{\max}} & > \ofrac{1-\epsilon}, \\
\frac{b(d_{\min}-2)}{2(1+\delta)c} & > \ofrac{1-\epsilon}.
\end{align*}
Recall that $t$ is the time from the start of the infection spreading to our observation of $G_n$. From Lemma \ref{lemma:N}, for each $\epsilon$, there exists $t_0$ such that if $t \geq t_0$, we have
\begin{align}
\frac{(d_{\min}-1)(1+\delta)N_{\min}(t)}{d_{\max}N_{\max}(t)} & > 1, \label{ineq:N1}\\
\frac{(d_{\min}-2)N_{\min}(t)}{2(1+\delta)N_{\max}(t)} & > 1. \label{ineq:N2}
\end{align}
We will make use of the two inequalities \eqref{ineq:N1} and \eqref{ineq:N2} extensively in the following proof steps. Let $\mathcal{E}_t$ be the event defined in Lemma \ref{lemma:N}. Then from Lemma \ref{lemma:N}, we have for $t \geq t_0$,
\begin{align}
\P(\tilde{S} = S^* \mid S^*) & \geq \P(\tilde{S} = S^*\mid S^*, \mathcal{E}_t)\P(\mathcal{E}_t \mid S^*) \geq (1-\epsilon)\P(\tilde{S} = S^*\mid S^*, \mathcal{E}_t). \label{ineq:Pbound}
\end{align}
In the following, we show that $\P(\tilde{S} = S\mid S, \mathcal{E}_t) = 1$ for $t \geq t_0$. The proof then follows from \eqref{ineq:Pbound} as $\epsilon$ can be chosen arbitrarily small.

\begin{figure}[!t] 
  \centering
  \psfrag{a}[][][0.8][0]{$s_1$}
  \psfrag{b}[][][0.8][0]{$s_2$}
  \psfrag{c}[][][0.8][0]{$x_{n-1}$}
  \psfrag{d}[][][0.8][0]{$x_1$}
  \psfrag{e}[][][0.8][0]{$u_0$}
  \psfrag{f}[][][0.8][0]{$u_1$}
  \psfrag{g}[][][0.8][0]{$u_m$}
  \psfrag{h}[][][0.8][0]{$w_1$}
  \psfrag{i}[][][0.8][0]{$w_k$}
  \psfrag{j}[][][0.8][0]{$v_0$}
  \psfrag{k}[][][0.8][0]{$v_1$}
  \psfrag{l}[][][0.8][0]{$v_l$}
  \psfrag{m}[][][0.8][0]{$y_1$}
  \psfrag{n}[][][0.8][0]{$y_{p-1}$}
  \includegraphics[width=0.5\textwidth]{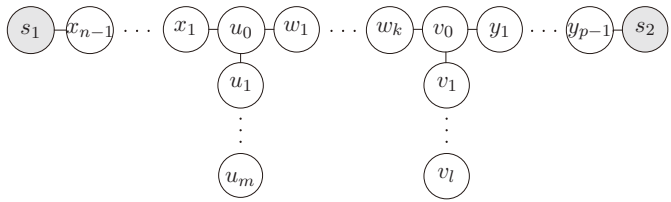}
  \caption{Illustration of the network structure when $u_0 \neq v_0$. Not all nodes are shown.}
  \label{fig:network_for_two_sources_case_1}
\end{figure}

\begin{figure}[!t] 
  \centering
  \psfrag{a}[][][0.8][0]{$s_1$}
  \psfrag{b}[][][0.8][0]{$s_2$}
  \psfrag{c}[][][0.8][0]{$w_0$}
  \psfrag{d}[][][0.8][0]{$w_1$}
  \psfrag{e}[][][0.8][0]{$w_2$}
  \psfrag{f}[][][0.8][0]{$w_k$}
  \psfrag{g}[][][0.8][0]{$v_1$}
  \psfrag{h}[][][0.8][0]{$v_2$}
  \psfrag{i}[][][0.8][0]{$v_l$}
  \psfrag{j}[][][0.8][0]{$u_1$}
  \psfrag{k}[][][0.8][0]{$u_2$}
  \psfrag{l}[][][0.8][0]{$u_m$}
  \includegraphics[width=0.5\textwidth]{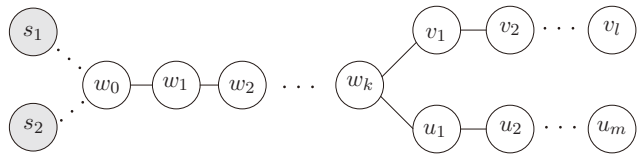}
  \caption{Illustration of the case where $u_0 = v_0 = w_0$.}
  \label{fig:network_for_two_sources_case_2}
\end{figure}

To show that $\P(\tilde{S} = S\mid S, \mathcal{E}_t) = 1$ is equivalent to showing that with probability one, $\hC{S}{G_n} > \hC{u_m,v_l}{G_n}$, for all node pairs $u_m,v_l \in G_n$ such that at least one of them is not in $S$. Let $u_0$ and $v_0$ be the first nodes in $\rho(s_1,s_2)$ that are connected to $u_m$ and $v_l$ respectively. We divide the proof into two cases, depending on whether $u_0$ and $v_0$ are distinct or not, as shown in Figures \ref{fig:network_for_two_sources_case_1} and \ref{fig:network_for_two_sources_case_2}.

Suppose that $u_0 \neq v_0$. A typical network for this case is shown in Figure \ref{fig:network_for_two_sources_case_1}, where $m,l,n,p$, and $k$ are non-negative integers, and at least one of $u_m$ and $v_l$ is not in $S$, i.e., either $m+l > 0$ or $n+p > 0$. We let $u_0 = s_1$ if $n=0$, and $v_0 = s_2$ if $p=0$.

We will show that if either $m+l > 0$ or $n+p > 0$, we have for $t\geq t_0$,
\begin{align}
\frac{\hC{s_1,s_2}{G_n}}{\hC{u_m,v_l}{G_n}} = \frac{\hC{s_1,s_2}{G_n}}{\hC{u_0,v_0}{G_n}}\cdot \frac{\hC{u_0,v_0}{G_n}}{\hC{u_m,v_l}{G_n}}> 1. \label{ineq:case1}
\end{align}
The proof follows by showing that $\hC{u_0,v_0}{G_n}\geq\hC{u_m,v_l}{G_n}$, where strict inequality holds if $m+l > 0$, and $\hC{s_1,s_2}{G_n}\geq\hC{u_0,v_0}{G_n}$ with strict inequality holding if $n+p > 0$. From \eqref{equ:approximation_permulation_count}, we have \footnote{We define products over empty sets to be 1.}
\begin{align*}
\frac{\hC{u_0,v_0}{G_n}}{\hC{u_m,v_l}{G_n}} & = \frac{Q(u_0 , v_0)}{Q(u_m , v_l)} \cdot \prod \limits_{w \in \rho (u_m , u_1) \cup \rho (v_l , v_1)} |T_w (u_0 , v_0)|^{-1} \\
& = [2(1+ \delta)]^{-(m+l)} \cdot \frac{\prod_{i=1}^{m+l+k+2} I_i^*(u_m, v_l)}{\prod_{i=1}^{k+2} I_i^*(u_0, v_0)} \cdot
\prod \limits_{w \in \rho (u_m , u_1) \cup \rho (v_l , v_1)}|T_w (u_0 , v_0)|^{-1}\\
& \geq [2(1+ \delta)]^{-(m+l)} \cdot \prod_{i=1}^{m+l} I_i^*(u_m, v_l) \cdot
\prod \limits_{w \in \rho (u_m , u_1) \cup \rho (v_l , v_1)}|T_w (u_0 , v_0)|^{-1}\\
& \geq \left[\frac{\max \{|T_{u_0}(u_m , v_l)|, |T_{v_0} (u_m , v_l)|\}}{2(1+ \delta)\cdot \max\left\{|T_{u_1} (u_0 , v_0)|,|T_{v_1} (u_0 , v_0)|\right\}}\right]^{m+l} \\
& \geq \left[ \frac{(d_{\max}-2)N_{\min}(t)+1}{2(1+\delta ) \cdot N_{\max}(t)} \right]^{m+l} \\
& > 1,
\end{align*}
if $m+l>0$. The first inequality follows because $I_{m+l+i}^*(u_m, v_l) \geq I_i^*(u_0, v_0)$ for $i=1,\ldots,k+2$, and the last inequality follows from \eqref{ineq:N2} when $t\geq t_0$.

Let $\psi = \deg_G(s_1)+\deg_G(s_1)$. We have for $t\geq t_0$,
\begin{align*}
\frac{\hC{s_1,s_2}{G_n}}{\hC{u_0,v_0}{G_n}} & = \frac{Q(s_1,s_2)}{Q(u_0 , v_0)} \cdot \prod \limits_{w \in \rho (s_1, x_1) \cup \rho (y_1 , s_2)} |T_w (u_0 , v_0)| \\
& = [2(1+ \delta)]^{n+p} \cdot \frac {\prod \limits_{i=1}^{k+2} I_i^*(u_0, v_0)}{\prod \limits_{i=1}^{n+p+k+2} I_i^*(s_1, s_2)} \cdot
\prod \limits_{w \in \rho (s_1, x_1) \cup \rho (y_1 , s_2)} |T_w (u_0 , v_0)|\\
& \geq [2(1+ \delta)]^{n+p} \cdot \prod \limits_{i=k+3}^{n+p+k+2} I_i^*(s_1, s_2)^{-1} \cdot \prod \limits_{w \in \rho (s_1, x_1) \cup \rho (y_1 , s_2)} |T_w (u_0 , v_0)|\\
& \geq \left[\frac{2(1+ \delta)\cdot \min \left\{|T_{s_1}(u_0,v_0)|,|T_{s_2}(u_0,v_0)|\right\}}{\psi N_{\max}(t)+2}\right]^{n+p} \\
& \geq \left[\frac {(1+ \delta)(d_{\min}-1) \cdot N_{\min}(t)+1+\delta}{d_{\max}N_{\max}(t)+1} \right]^{n+p}\\
& > 1,
\end{align*}
where the first inequality follows because $I_{i}^*(u_0, v_0) \geq I_i^*(s_1, s_2)$ for $i=1,\ldots,k+2$, and the last inequality follows from \eqref{ineq:N1} if $n+p>0$. The bound \eqref{ineq:case1} is now proved.

We next consider the case where $u_0 = v_0 = w_0$ in Figure \ref{fig:network_for_two_sources_case_2}, where $k,m$ and $l$ are non-negative integers. When $t\geq t_0$, we have the following bounds, which are straight forward to verify and whose proofs are omitted here.
\begin{enumerate}[(i)]
\item $I_i^*(u_m,v_l) \geq (\psi-2)N_{\min}(t) + 2 \geq (d_{\min}-2)N_{\min}(t)$ for $i=1,\ldots,d(u_m,v_l)+1$,
\item $I_i^*(s_1,s_2) \leq \psi N_{\max}(t)+2 \leq 2d_{\max} N_{\max}(t) + 2$ for all $i=1,\ldots,d(s_1,s_2)+1$,
\item $|T_{w_i} (u_m , v_l)| \geq (\psi-2)N_{\min}(t) + 2 \geq (d_{\min}-2)N_{\min}(t)$ for all $i=1,\ldots,k-1$,
\item $|T_{w} (u_m , v_l)| \geq (d_{\min}-1)N_{\min}(t) + 1$ for all $w\in \rho(s_1,s_2)$,
\item $|T_{w_i} (s_1,s_2)| \leq N_{\max}(t)$ for all $i=1,\ldots,k-1$, and
\item $|T_{w} (s_1,s_2)| \leq N_{\max}(t)$ for all $w\in \rho(u_m,v_l)$.
\end{enumerate}
The above bounds yield
\begin{align*}
& \frac{\hC{s_1,s_2}{G_n}}{\hC{u_m,v_l}{G_n}} \\
 =&  \frac{Q(s_1,s_2)}{Q(u_m , v_l)} \frac{\prod_{w \in G_n \backslash \rho (u_m, v_l)} |T_w (u_m , v_l)|}{\prod_{w \in G_n \backslash \rho (s_1,s_2)} |T_w (s_1,s_2)|}  \\
 =& (2(1+\delta))^{d(s_1,s_2)-d(u_m , v_l)} \frac{\prod_{i=1}^{d(u_m, v_l)+1} I_i^*(u_m,v_l)}{\prod_{i=1}^{d(s_1,s_2)+1} I_i^*(s_1,s_2)} \frac{\prod_{i=1}^{k-1}|T_{w_i} (u_m , v_l)| \prod_{w \in \rho (s_1,s_2)} |T_w (u_m , v_l)|}{\prod_{i=1}^{k-1}|T_{w_i} (s_1,s_2)| \prod_{w \in \rho (u_m , v_l)} |T_w (s_1,s_2)|} \\
=& \prod_{i=1}^{k-1}\frac{|T_{w_i} (u_m , v_l)|}{|T_{w_i} (s_1,s_2)|} \cdot (2(1+\delta))^{-d(u_m , v_l)-1}\frac{\prod_{i=1}^{d(u_m, v_l)+1} I_i^*(u_m,v_l)}{\prod_{w \in \rho(u_m , v_l)} |T_w (s_1,s_2)|} \cdot (2(1+\delta))^{d(s_1,s_2)+1}\frac{\prod_{w \in \rho (s_1,s_2)}|T_w (u_m , v_l)|}{\prod_{i=1}^{d(s_1,s_2)+1} I_i^*(s_1,s_2)} \\
\geq & \left[ \frac{(d_{\min}-2)N_{\min}(t)}{N_{\max}(t)} \right]^{k-1} \left[ \frac{(d_{\min}-2)N_{\min}(t)}{2(1+\delta)N_{\max}(t)} \right]^{d(u_m,v_l)+1} \left[ \frac{(1+\delta)((d_{\min}-1) N_{\min}(t)+1)}{d_{\max} N_{\max}(t)+1}\right]^{d(s_1,s_2)+1} \\ > & 1,\end{align*}
where the last inequality follows from \eqref{ineq:N1} and \eqref{ineq:N2}. The theorem is now proved.

\section{Proof of Theorem \ref{theorem:TSE_differentiate}}\label{appendix:theorem:TSE_differentiate}
Let $t$ be the elapsed time from the start of an infection spreading from a single $s$ to the time we observe $G_n$. We wish to show that Algorithm TSE estimates as sources $s$ and one of its neighbors with probability (conditioned on $s$ being the infection source) converging to $1$ as $t\to \infty$. This is equivalent to showing that for $t$ sufficiently large, and for each pair of nodes $u_m,v_l \in G_n$ where either $d(u_m,s) > 1$ or $d(v_l,s) > 1$, there exists a neighbor $r$ of $s$ such that $\hC{s,r}{G_n} > \hC{u_m,v_l}{G_n}$.

\begin{figure}[!t] 
  \centering
  \psfrag{a}[][][0.8][0]{$$}
  \psfrag{b}[][][0.8][0]{$r$}
  \psfrag{c}[][][0.8][0]{$s$}
  \psfrag{d}[][][0.8][0]{$w_1$}
  \psfrag{e}[][][0.8][0]{$w_k$}
  \psfrag{f}[][][0.8][0]{$v_1$}
  \psfrag{g}[][][0.8][0]{$v_2$}
  \psfrag{h}[][][0.8][0]{$v_l$}
  \psfrag{i}[][][0.8][0]{$u_l$}
  \psfrag{j}[][][0.8][0]{$u_2$}
  \psfrag{k}[][][0.8][0]{$u_m$}
  \includegraphics[width=0.4\textwidth]{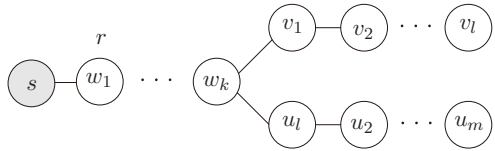}
  \caption{A typical network for a single source tree.}
  \label{fig:network_for_one_source_case_2}
\end{figure}

A typical network is shown in Figure \ref{fig:network_for_one_source_case_2}, where $k,m$ and $l$ are non-negative integers. If $m,l$ and $k$ are positive, we let $r$ be the neighbor of $s$ that lies on the path connecting $s$ to $u_m$ (i.e., the node $w_1$ in Figure \ref{fig:network_for_one_source_case_2}). If $m$ and $l$ are positive and $k=0$, then $r$ is chosen to be either $u_1$ or $v_1$. If $m=0$, we must have $k>0$ so that $w_k = u_m$ and $r = w_1$. A similar remark applies for the case $l=0$. Note that $m+l > 0$. For $t$ sufficiently large, we have
\begin{align*}
\frac{\hC{s,r}{G_n}}{\hC{u_m,v_l}{G_n}} & = \frac{Q(s,r)}{Q(u_m,v_l)} \cdot \frac{\prod \limits_{w \in G_n \backslash \rho (u_m,v_l)}|T_w(u_m, v_l)|}{\prod \limits_{w \in G_n \backslash \{s,r\}}|T_w(s,r)|}\\
&=[2(1+\delta)]^{1-(m+l)} \cdot \frac{\prod_{i=1}^{m+l+1} I_i^*(u_m,v_l)}{\prod_{i=1}^{2} I_i^*(s,r)}\cdot  \frac{\prod_{w\in\rho(s,w_{k-1})} |T_{w}(u_m,v_l)|}{\prod_{i=2}^{k-1} |T_{w_i}(s,r)| \cdot \prod \limits_{w \in \rho (u_m,v_l)}|T_w(s,r)|}\\
&=[2(1+\delta)]^{1-(m+l)} \cdot \prod_{i=1}^{m+l} I_i^*(u_m,v_l)\cdot \frac{\prod_{i=1}^{k-1} |T_{w_i}(u_m,v_l)|}{\prod_{i=2}^{k-1} |T_{w_i}(s,r)| \cdot \prod \limits_{w \in \rho (u_m,v_l)}|T_w(s,r)|}\\
&\geq [2(1+\delta)]^{1-(m+l)} \cdot |T_{w_k}(u_m,v_l)|^{m+l}\cdot \frac{|T_{s}(u_m,v_l)|^{k-1}}{N_{\max}(t)^{k-2} \cdot N_{\max}(t)^{m+l+1}}\\
& \geq \left[2(1+\delta)\right]^{k} \cdot \left[\frac{(d_{\min}-1)N_{\min}(t)}{2(1+\delta)\cdot N_{\max}(t)}\right]^{m+l+k-1} \\
& > 1,
\end{align*}
where the last inequality follows from \eqref{ineq:N2} and Lemma \ref{lemma:N}. The proof of the theorem is now complete.

\bibliography{IEEEabrv,WQ}{}
\bibliographystyle{IEEEtran}
\end{document}